\documentclass[reprint,amsmath,amssymb,pra,aps,nofootinbib,superscriptaddress,floatfix]{revtex4-1}

\usepackage[T1]{fontenc}
\usepackage[utf8]{inputenc}
\usepackage{amsmath, amssymb}
\usepackage{bbold}
\usepackage{bm}
\usepackage{tipa}
\usepackage{subfigure}
\usepackage{blindtext}
\usepackage[dvipsnames]{xcolor}
\definecolor{dcyan}{RGB}{0,100,100}
\usepackage{graphicx}
\usepackage{tabularx}
\usepackage{textgreek}
\usepackage{xfrac}
\usepackage{units}
\usepackage{comment}
\usepackage{float}
\usepackage{placeins}
\usepackage{upgreek}
\usepackage{booktabs}  
\usepackage{enumitem}  

\usepackage{pifont}
\usepackage{xcolor}
\usepackage{xparse}
\usepackage{tikz}

\usepackage[hypertexnames=false]{hyperref}
\usepackage{xcolor}
\definecolor{green_cust}{RGB}{0,154,85}
\definecolor{red_cust}{RGB}{173,49,54}
\definecolor{blue_cust}{RGB}{0,103,148}
\hypersetup{colorlinks=true, linkcolor=red_cust, citecolor=green_cust, filecolor=blue_cust, urlcolor=blue_cust}

\makeatletter

\renewcommand\onecolumngrid{
\do@columngrid{one}{\@ne}%
\def\set@footnotewidth{\onecolumngrid}
\def\footnoterule{\kern-6pt\hrule width 1.5in\kern6pt}%
}

\renewcommand\twocolumngrid{
        \def\footnoterule{
        \dimen@\skip\footins\divide\dimen@\thr@@
        \kern-\dimen@\hrule width.5in\kern\dimen@}
        \do@columngrid{mlt}{\tw@}
}%

\makeatother

\usepackage{soul}

\newcommand{\Applied}{Department of Applied Physics, Stanford University, Stanford, CA}
\newcommand{\Physics}{Department of Physics, Stanford University, Stanford, CA}

\definecolor{reviewcolor}{HTML}{9381FF}
\newcommand{\review}[1]{#1}

\newcommand{\bra}[1]{\langle{#1}|}
\newcommand{\ket}[1]{|{#1}\rangle}

\newcommand{\Figref}[1]{Fig.~\hyperref[#1]{\ref{#1}}}

\begin{document}
\title{A 10 Megahertz Spatial Light Modulator}

\author{Xin Wei}\thanks{These authors contributed equally.}
\affiliation{\Applied}
\author{Zeyang Li}\thanks{These authors contributed equally.}
\affiliation{\Applied}
\affiliation{\Physics}
\author{Abhishek V. Karve}
\affiliation{\Applied}
\author{Adam L. Shaw}
\affiliation{\Applied}
\affiliation{\Physics}
\author{David I. Schuster}
\affiliation{\Applied}
\author{Jonathan Simon}\email{jonsimon@stanford.edu}
\affiliation{\Applied}
\affiliation{\Physics}


\def\ripa{Re-Imaging Phased Array}

\begin{abstract}
Rapid and programmable shaping of light fields is central to modern microscopy~\cite{Hell2007,Vettenburg2014,fahrbach2010microscopy}, display technologies, optical communications and sensing~\cite{takeda1982fourier,richardson2013space,bozinovic2013terabit}, quantum engineering~\cite{steinhauer2016observation,clark2017collective,Chiu2018,Yan2023,Yamazaki2010,baum2022optical,Orsi2024,marsh2025multimode}, and quantum information processing~\cite{graham2022multi,radnaev2025universal,manetsch2025tweezer,endres2016atom,lam2021demonstration,chiu2025continuous,li2025fastcontinuouscoherentatom,Chow2024,bluvstein2024logical,guo2025towards}.
Current wavefront shaping technologies face a fundamental dichotomy: \review{liquid-crystal-on-silicon} spatial light modulators (\review{LCoS-}SLMs) offer high pixel count but suffer from low refresh rates, while acousto-optic deflectors (AODs) provide moderate speed with restricted optical beam geometries~\cite{brown2019gray,bluvstein2022quantum}. Though recent advances in photonic integrated circuits achieve fast switching~\cite{zhao2025integrated}, there is currently no tool that provides MHz-rate, continuous motion, and arbitrarily reconfigurable control over a set of diffraction-limited spots.
Here we introduce a new class of spatial light modulator that provides both 2D pixel geometry and high speed.
The device operates by encoding spatial information in frequency bins via a broadband optical phase modulator, and decoding them via a first-of-its-kind, high-resolution 2D spectrometer.
The spectrometer, based on the architecture which we call the \ripa~(RIPA), achieves its sensitivity through long path-lengths, enabled by intra-spectrometer re-imaging lens-guides.
We demonstrate site-resolved optical pulsing with a 44(1)~ns rise time, corresponding to frame rates exceeding 10 million frames per second, as well as arbitrary, reconfigurable 2D addressing and multi-site operations, including asynchronous, independent beam motion, splitting, and recombination. 
Leveraging these tools opens new horizons in rapid optical manipulation of matter across science, from fast, scalable control that approaches the inertial and radiation limits of atoms in quantum processors, to dynamically programmable, microsecond-resolved illumination in microscopy and neuro-biological imaging~\cite{katona2012fast,yamaguchi2023multi}.

\end{abstract}
\date{\today}
\maketitle

High-speed spatial light modulation enables diverse applications across disparate length scales and technical fields, from macroscopic 3D optical metrology~\cite{takeda1982fourier}, to microscopic wave engineering for high-contrast imaging~\cite{Hell2007,Vettenburg2014,fahrbach2010microscopy}, as well as optically multiplexed communication~\cite{richardson2013space,bozinovic2013terabit}.
In atomic, molecular and optical physics, it enables dynamically switchable optical traps~\cite{steinhauer2016observation,clark2017collective,Chiu2018}, tunable coupling strength in quantum optical systems~\cite{Yan2023}, global and local modulation in quantum gases~\cite{Yamazaki2010,clark2017collective}, and arbitrary excitation in multi-mode cavity quantum electrodynamics~\cite{baum2022optical,Orsi2024,marsh2025multimode}.

For neutral-atom quantum computers in particular, high speed local addressing tools are critical for \review{independent} atom control. Increasing the speed of these operations while ensuring they remain independently actuated has the potential to accelerate numerous computing primitives, from faster \review{\textit{in-situ} gate operations}~\cite{graham2022multi,radnaev2025universal}, to independent atom movement~\cite{manetsch2025tweezer,endres2016atom,lam2021demonstration} 
which enables continuous atom reloading~\cite{chiu2025continuous,li2025fastcontinuouscoherentatom,Chow2024}, \review{velocity-selective addressing}~\cite{lib2026velocity} \review{and other flying-qubit architectures}~\cite{xue2026factoring2048bitrsa,dudinets2025all}, \review{plus} more efficient execution of quantum algorithms by relaxing the transversal-block-gate \review{co-design}~\cite{eastin2009restrictions,bluvstein2024logical,pogorelov2025experimental,guo2025towards,bravyi2024high} necessary for high-rate computation.

Existing local addressing solutions offer either high pixel count or high frame rate, but not both: Digital micromirror devices (DMDs) and conventional \review{liquid-crystal-on-silicon} spatial light modulators (\review{LCoS}-SLMs) both offer 2D image generation but are limited to $\sim$kilohertz frame rates by their physics: either mirror inertia or the liquid-crystal response time.
Combining a DMD and an SLM has recently demonstrated large-scale 2D modulation at frame rates up to 43~kHz~\cite{zhang2024scaled}, which alleviates challenges in local gates, but is insufficient for independent atom transport.
Crossed acousto-optic deflectors (AODs) provide microsecond\review{-level} switching and beam steering, yet intrinsically generate either a single spot, or a grid of spots (an outer product of one-dimensional patterns)~\cite{brown2019gray,bluvstein2022quantum}, limiting independent addressing capabilities. 
Recent approaches based on integrated electro-optics deliver impressive bandwidth~\cite{panuski2022full,Menssen2023Scalable,zhao2025integrated}, but have yet to be demonstrated at scale.

We introduce a high-bandwidth spatial light \review{modulator} that fundamentally departs from these conventional approaches. 
Current technologies rely on piecing together many low-bandwidth modulators (e.g., liquid-crystal, micro-electromechanical systems (MEMS)) and stacking parallel control lines to increase aggregate bandwidth.
\review{We instead employ one extremely fast $\sim 10$ GHz electro-optical modulator (EOM) to encode each pixel of an image into a distinct frequency bin of a single optical channel, and then map that frequency to the correct location on a display using a high-resolution spectrometer.}
Achieving high pixel count requires many narrowly-spaced frequency-bins, and converting these frequency bins to spatial locations thus necessitates a dispersive element with exceptionally high spectral resolution.
Conventional dispersive devices face intrinsic limitations: diffraction gratings are limited by physical beam size, while Virtually Imaged Phased Arrays (VIPAs)~\cite{shirasaki_VIPA,chan20082,sadiek2024airspaced}---which enhance spectral dispersion by circulating the beam within an etalon---suffer from diffractive beam expansion that degrades spatial mode overlap. 
Similarly, optical cavities offer high dispersion but only spatially resolve a single output frequency.

Motivated by these considerations, we demonstrate \review{a spatial light modulator} using self-imaging VIPAs---which we call \ripa~(RIPAs)---that enable a high-speed display via programmable, frequency-encoded modulation.
\review{Its} update rates exceeds 10~MHz, two orders of magnitude faster than AODs or the recent combined DMD-\review{LCoS-SLM} \review{device}~\cite{zhang2024scaled}. In what follows, we first outline the operating principle of the \review{RIPA-SLM}, then characterize a simplified 1D RIPA, extend it to 2D to demonstrate arbitrary pattern generation, and quantify its performance. Finally, we demonstrate sub-microsecond operation, directly confirming its ability to perform high-speed, multiplexed site addressing.

\begin{figure*}[!hbtp]
	\centering
 	\includegraphics[width=183mm]{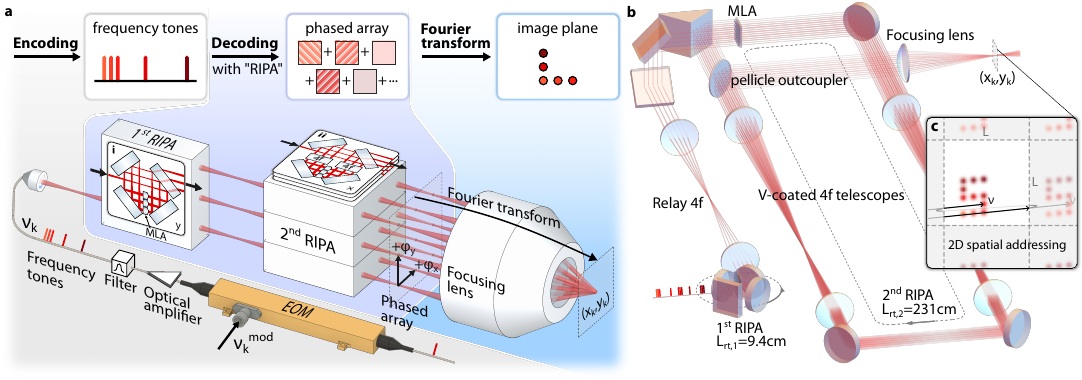}
	\caption{
        \textbf{Operation of a 10 Mega-FPS spatial light modulator.}
        \textbf{a,} \review{The \ripa - spatial light modulator (RIPA-SLM) enables high-bandwidth spatial addressing by \emph{encoding} the spatial addressing pattern into frequency tones and \emph{decoding} them to phased arrays with a frequency-dependent phase gradient, which then \emph{Fourier transforms} to distinct spots in the image plane.} 
        A coherent input beam is phase-modulated by a high-bandwidth electro-optic modulator (EOM) to generate designed frequency tones $\nu_k$ (shades of red). \review{An optical amplifier is used to boost the optical power after the modulator.}
        The modulated beam is converted into a 2D beam array by the first and second RIPAs (insets \textbf{i} and \textbf{ii}). Each RIPA duplicates and displaces the beam over each round trip, during which the beams acquire relative propagation phases $\varphi_y$($\varphi_x$), determined by the round-trip path lengths of the first (second) RIPA and the laser frequency.
        Beam diffraction over the long round trips is suppressed by refocusing optics, yielding a uniform transverse mode across the 2D phased array. 
        Interfering these beams with a focusing lens performs frequency-to-position mapping. 
        \textbf{b,} Detailed optical layout of the RIPA system. 
        Diffraction-less propagation is accomplished using a running-wave pseudo-cavity geometry with: an intra-path microlens array (MLA) in the first RIPA, and an identical MLA + two 4f telescopes in the second RIPA (to add propagation distance without spatial inversion). \review{Quantitatively, the RIPAs have a round-trip path-length of} $L_{\mathrm{rt,1}}=9.4~\mathrm{cm}$ and $L_{\mathrm{rt,2}}=231\mathrm{cm}$, respectively (drawn to scale here).
        \textbf{c,} The RIPA-SLM directs each tone (shades of red) to a specific location $(x_k,y_k)$.
        Increasing the frequency rasters the spot through the Brillouin zone, wrapping around at the edge. 
        Programmable RF modulation therefore synthesizes arbitrary 2D images at the spectrometer output.
        }
	\label{fig:fig1}
\end{figure*}

\section*{Principle of Operation}

\review{At a high level, our \ripa- spatial light modulator (RIPA-SLM) operates in three steps, as shown in Fig.~\ref{fig:fig1}\textbf{a}.
First, an electro-optic modulator (EOM) encodes the desired spatial pattern into a set of optical frequency tones via RF sideband modulation, with each optical tone carrying the field for one location in the spatial pattern.
Second, two cascaded RIPA devices decode each optical tone into a coherent 2D \emph{phased array} with a frequency-dependent phase gradient.
Third, a focusing lens performs a Fourier transform of this phased array, mapping each optical tone to its corresponding location in the image plane.}

\begin{figure}[!ht]
    \centering
    \includegraphics[width=89mm]{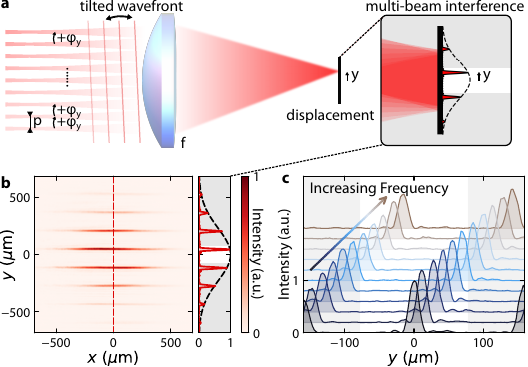}
    
    \caption{\textbf{1D phased-array frequency-to-position mapping.} 
    For a single frequency \review{linearly polarized} input, a 1D RIPA produces, at its output \textbf{a}, an array of parallel beams with pitch $p$ and a controllable phase step between adjacent beams $\varphi_y$, \review{which resembles a tilted plane-wavefront. A lens with focal length $f$ interferes these beams to produce an intensity pattern whose peak position $y$ is set by $\varphi_y$. 
    Because the phased array consists of many discrete beamlets, the same constructive-interference condition occurs at multiple diffraction orders, producing the multi-peaked structure shown in the \textbf{inset},
    with an envelope (dashed curve) set by the single-beam size; within the first Brillouin zone (unshaded), the pattern contains a single diffraction-limited peak.}
    \textbf{b}, Measured focal-plane intensity distribution for a single-frequency input. The dashed vertical line indicates the line-cut shown to the right, where the red curve is the measured line-cut profile and the dashed black curve is the Gaussian envelope corresponding to a single beam in the phased array.  
    \textbf{c}, Intensity profiles along the line-cut in \textbf{b} for increasing laser frequency, showing continuous peak translation across the image plane consistent with the designed frequency-to-position mapping.}
    \label{fig:fig2}
\end{figure}

\review{A key concept underlying the operation of our spatial light modulator is the \emph{phased array}, which consists of an ensemble of coherent optical beams with ordered spatial positions and a well-defined phase relationship between them. If adjacent beams have phase differences $\varphi_{x}(\varphi_{y})$ in the $x$($y$) directions, then the combined field arising from the interference of these beams acquires propagation-direction tilts along $x$($y$) that are proportional to $\varphi_{x}(\varphi_y)$. This may be understood from the fact that a phased array with a constant phase difference between adjacent beams forms a discrete analogue of a tilted planar wavefront (see Fig.~\ref{fig:fig2}\textbf{a}).}

\review{
Starting from a single-frequency laser beam input, the simplest way to generate a 1D phased array is to repeatedly circulate the beam while out-coupling a small fraction of it on each pass.
By doing so, each out-coupled beam picks up a time delay $T_y$ compared to its immediate predecessor.
In this way, the delay of the $j^{\mathrm{th}}$ out-coupled beam in 1D is $j\times T_y$, and for a fixed laser frequency $\nu$, it acquires a phase $\varphi_{1,j}=j\times T_y \times(2\pi\nu)\equiv j\times \varphi_{y}$. Increasing the frequency $\nu$ of the input laser beam thus increases the phase step $\varphi_y$ between adjacent beams in the phased array, monotonically deflecting the direction of constructive-interference until adjacent beams acquire a phase difference of $2\pi$, at which point the wavefront hops back to its initial direction~\cite{von1960properties}.}

\review{We realize this delay-and-out-couple architecture, termed \ripa~(RIPA), as shown in Fig.~\ref{fig:fig1}\textbf{a}, inset \textbf{i}: A single Gaussian beam enters and \review{circulates repeatedly} within a four-mirror loop aligned to introduce a small transverse offset on each round trip. A partially reflecting mirror out-couples a fraction of the beam on each round trip, creating a 1D phased array. 
We place a microlens array (MLA) so that each round trip passes through a different microlens. Each microlens \emph{re-images} the spatial mode (hence the \emph{Re-Imaging} in RIPA) onto itself up to a transverse displacement, thereby combating beam divergence and stabilizing the transverse mode by forming a lens-guide centered on the beam path (see Methods).
Thus, the RIPA broadcasts the input beam into a 1D phased array of its copies with identical transverse modes and a well-defined phase increment between adjacent beams.}

To extend this approach to a 2D phased array, we cascade a \review{small-delay RIPA along $y$ and a large-delay RIPA along $x$, as illustrated in Fig.~\ref{fig:fig1}\textbf{a}, insets \textbf{i} and \textbf{ii}.
After propagating through both RIPAs, the $(i,j)^{\mathrm{th}}$ beam experiences a time delay of $T_{i,j}=i\times T_x + j\times T_y$, with $T_x$($T_y$) being the round-trip delay of the RIPA along $x$($y$).
This establishes a 2D phased array with a uniform spatial profile and a linear phase profile $\varphi_{i,j} = i\times \varphi_x + j\times \varphi_y$, resembling a \emph{tilted planar} wavefront whose tilt is controlled by the input laser frequency $\nu$.}
Focusing this frequency-controlled phased array through a lens into an image plane performs the Fourier transform of the array, thereby mapping the tilted wavefront to a diffraction-limited spot at $(x,y)$.
A large ratio between the $x$ and $y$ delays ensures that this \emph{frequency-to-position mapping} rasters repeatedly across $x$, while slowly deflecting along $y$. The full frequency spectrum is thus \review{mapped \emph{row by row} onto the image plane}, (see Fig.~\ref{fig:fig1}\textbf{c}), akin to the raster-pattern of a cathode ray tube (CRT).

\review{
Achieving a frequency resolution $\delta \nu\sim$ 10 MHz is necessary to access a large number of resolvable points within the $\sim$ 10 GHz bandwidth of our modulator. This requirement, however, necessitates a tens-of-meter-scale optical path length in the large-delay RIPA, which would require unreasonably large microlenses to refocus the diverging beams.
We instead combat this divergence by introducing in-loop 4F relay systems whose aberrations are conveniently suppressed by the existing microlens array lens-guide (see Methods). The RIPA round-trip phase ($\varphi_x$ and $\varphi_y$) is sensitive to path-length drift, which we actively suppress using a single piezo-translated mirror (see Methods).
}

By simultaneously injecting multiple optical tones at frequencies $\{\nu_k\}$ into the cascaded RIPAs, the output forms a sum of phased arrays with different characteristic phases $\{(\varphi_{x,k},\varphi_{y,k})\}$; the focused pattern thus simultaneously addresses multiple locations $\{(x_k,y_k)\}$ in the image plane.
Generating these tones via a high-bandwidth electro-optic modulator \review{with subsequent spectral filtering}~\cite{Li2025Filter} (see Methods) allows independent amplitude control to produce arbitrary intensity patterns (Fig.~\ref{fig:fig1}\textbf{c}).
\review{Despite the limited power handling of typical high-bandwidth EOMs, we can inject the spatially single-mode output of the modulator into a high-power optical amplifier, providing essentially unlimited power-handling capabilities for the RIPA-SLM (see Methods).}

\section*{Single-RIPA Interferometer Characterization}

\begin{figure*}[!htbp]
    \centering
    \includegraphics[width=\textwidth]{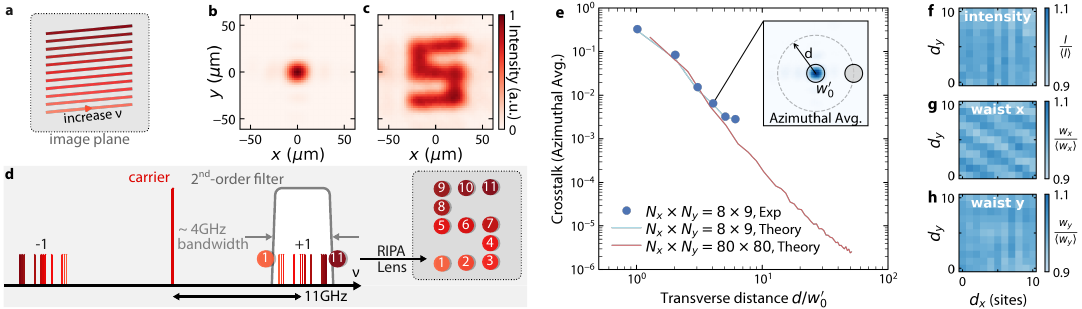}

    \caption{\textbf{Spectrally programmed 2D image synthesis and static characterization.}
    \textbf{a}, Principle of 2D frequency-to-position mapping: Sweeping the laser frequency $\nu$ (color scale) drives the addressing spot along a tilted raster trajectory across the image plane row-by-row.
    \textbf{b}, Injection of a single optical tone generates a (measured) intensity distribution corresponding to a diffraction-limited spot.
    \textbf{c}, Synthesis of an "S"-shaped pattern via simultaneous injection of multiple optical tones.
    \textbf{d}, Frequency-domain synthesis of the pattern in \textbf{c}. The pattern can be decomposed into 11 independent spots, each corresponding to a unique optical frequency $\nu_k$ that the RIPA+lens system maps to a distinct spatial location. This optical spectrum is generated by RF driving of a high-bandwidth electro-optic phase modulator, which in practice produces symmetric optical sidebands on the input light and leaves a strong carrier. A custom-built, second-order flat-top optical filter~\cite{Li2025Filter} transmits only the +1 order blue sidebands, suppressing all other orders and the carrier.
    \textbf{e}, \review{The crosstalk in this system exhibits power-law scaling that arises directly from the discreteness of the phased array.} Here we plot the measured (blue circles) and predicted (blue curve) crosstalk between two addressed sites as a function of their \review{transverse} separation $d$, in units of the $1/e^2$ intensity radius $w_0^\prime$. Scaling to a \review{$80\times 80$} array (red curve) predicts crosstalk levels below $10^{-4}$ at \review{a separation of 13 waists}.
    \textbf{Inset}: Crosstalk is calculated \review{in the image plane by azimuthally averaging} (along the dashed circle) the integrated intensity within one beam waist radius (solid circle). 
    \textbf{f–h} System-wide uniformity maps across the first Brillouin zone illustrating \textbf{f}, peak intensity and \textbf{g,h}, fitted beam waists $w_x$ ($w_y$) in the $x$- ($y$-) directions. The system exhibits high uniformity, with relative standard deviations ($\sigma$) of $2.6\%$ for peak intensity and $3.1\%$ ($1.9\%$) for the $w_x$ ($w_y$).}
    \label{fig:fig3}
\end{figure*}

To verify the concept of generating a frequency-encoded phased array, we characterize a single RIPA by examining the interference pattern of its coherent 1D phased array output, as shown in Fig.~\ref{fig:fig2}\textbf{a}.
The beams exiting this RIPA (before the lens in Fig.~\ref{fig:fig2}\textbf{a}) are parallel, \review{evenly spaced (with pitch $p$) Gaussian beams with waist $w_0$, and a constant phase step $\varphi_y$ between adjacent beams.}
\review{Thus, they resemble a tilted planar wavefront and when focused, constructively interfere to form a peak whose displacement $y$ depends linearly on $\varphi_y$.}
\review{In practice, this constructive-interference condition occurs at multiple diffraction orders; consequently, the} resulting pattern (insets in Fig.~\ref{fig:fig2}\textbf{a}) consists of a periodic grid of such peaks across several \emph{Brillouin zones} (in analogy with band structures in solid-state physics, see Methods).
Depending on the application, the ratio between the pitch $p$ and the beam waist $w_0$ can be tuned (e.g., using an MLA telescope, see Methods) to either distribute the power across multiple diffraction orders ($p\gg w_0$) or concentrate the power within a single order ($p\gtrsim w_0$).
Moving forward, we will only consider the light in the lowest Brillouin zone \review{(unshaded region in Fig.~\ref{fig:fig2}\textbf{a-c})}. 

As illustrated in Fig.~\ref{fig:fig2}\textbf{b}, the interference of the phased array (\review{containing $N_y=9$ beams}) produces a central feature whose $1/e^2$ intensity radius is reduced from the width of the Gaussian envelope, $w_{\mathrm{env}}=\frac{\lambda f}{\pi w_0}=460~\mu$m, to $w_0^\prime=\frac{\lambda f}{\pi p}\sqrt{\frac{6}{(N_y^2-1)}}=14~\mu$m (see Methods). \review{We choose the focal length to be $f=20$~cm here for camera-based characterization, so that the focused feature spans several camera pixels~\cite{zhang2024scaled}} (see Methods). The diffraction peaks (in higher Brillouin zones) are separated by $L=\frac{\lambda f}{p}=156~\mu$m, consistent with the MLA pitch $p=1~$mm. 

\review{As explained in the previous section, increasing the laser frequency $\nu$ increases the phase step $\varphi_y$ between beams and monotonically deflects the constructive-interference direction.}
We probe this frequency response in Fig.~\ref{fig:fig2}\textbf{c}.
The interference peak position $y$ shifts linearly with optical frequency $\nu$, governed by $y=\frac{\Delta \nu}{\mathrm{FSR}_1}L$. Here, the frequency shift $\Delta\nu$ is defined relative to the frequency where the \review{phase step} $\varphi_y$ satisfies $\varphi_y \equiv 0 \pmod{2\pi}$.
The $\mathrm{FSR}_1=\frac{c}{L_{\mathrm{rt,1}}}=3.19$~GHz is the free spectral range (FSR) of the first RIPA ($L_{\mathrm{rt,1}}=9.4\text{cm}$, see Fig.~\ref{fig:fig1}\textbf{b}), and equivalently, the total spectral bandwidth of the light exiting the RIPA-SLM.

\section*{Cascaded-RIPA Modulator Characterization}
\begin{figure*}[!htbp]
	\centering
 	\includegraphics[width=183mm]{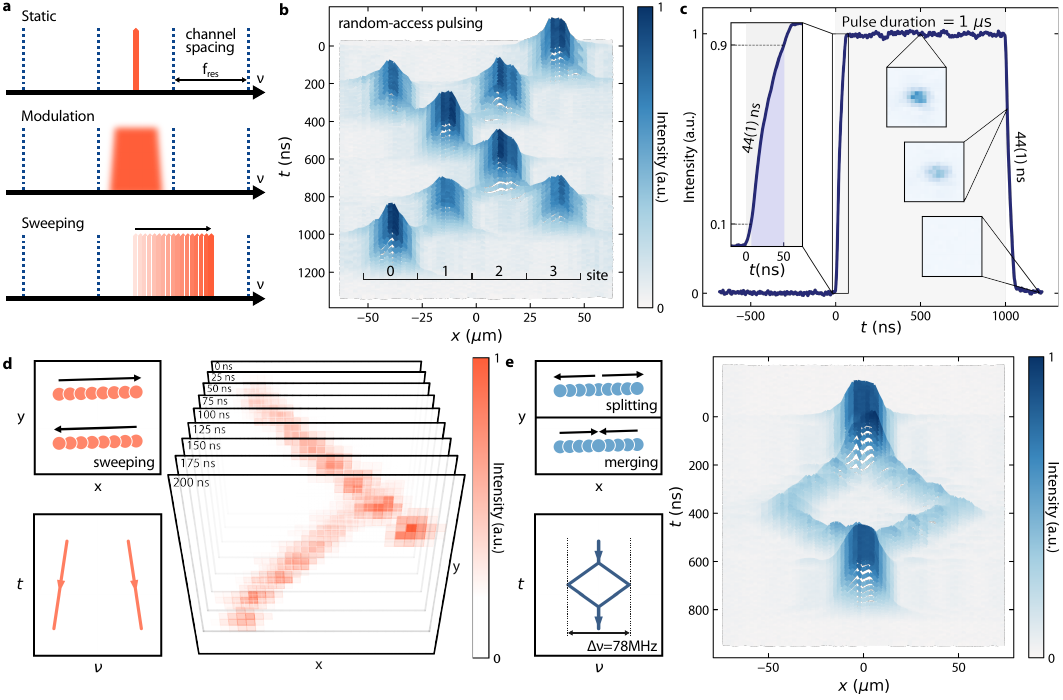}
    
    \caption{\textbf{Dynamic operation of the RIPA spatial light modulator and time-resolved characterization.}
    \textbf{a}, Operational regimes of the RIPA-SLM: \emph{static} site addressing with channel spacing $f_{\mathrm{res}}$; \emph{modulation} in amplitude (phase) which broadens the spectrum up to the channel spacing; and \emph{sweeping} frequency to move the addressing site continuously in space.
    \textbf{b}, Fast random-access addressing across a four-site registry demonstrating independent amplitude modulation of individual SLM pixels. The plot displays vertically stacked line-cuts sampled every $4~\text{ns}$ using a scanning photodetector (PD).
    \textbf{c}, Temporal response of a $1~\mu\text{s}$ optical pulse demonstrating symmetric 44(1)~ns rise and fall times ($10\%$–$90\%$). The left inset zooms in on the rising edge. The three central insets depict 2D intensity distributions reconstructed from PD scans in the fully ``on'' state (top), during the transient (middle), and in the ``off'' state (bottom).  
    \textbf{d}, Independent high-speed control of two spatially separated spots demonstrating continuous motion. Each spot corresponds to a distinct RF tone; independent linear frequency ramps applied to these tones are mapped by the RIPA to the continuous, independent motion of the respective spots in the image plane (insets). The main panel shows the spatiotemporal dynamics of two spots sweeping across the image plane at discrete time slices spanning $\Delta t = 200~\text{ns}$. 
    \textbf{e}, Programmable splitting and merging of a single spot. Starting from a single addressed spot, the RF drive is shaped to split the spot into two, translate each, and then recombine, as encoded by the frequency trajectories in the inset.}
	\label{fig:fig4}
\end{figure*} 

To form a 2D pattern, we direct the 1D phased array output of the first RIPA into the second RIPA \review{with a unit-magnification relay}, as shown in Fig.~\ref{fig:fig1}\textbf{b}.
\review{Here we choose to use identical MLAs in both RIPAs. The round-trip propagation matrices of the two RIPAs are thus identical (but not unity), such that the output of the first RIPA is mode-matched for injection into the second RIPA.}
The cascaded system generates a 2D $N_x\times N_y$ phased array with a phase profile of $\varphi_{i,j}=i\times\varphi_x+j\times\varphi_y$.
As in the 1D case, all beams within the array share a common transverse mode profile, and their locations remain \review{ordered and} invariant under changes in laser frequency. \review{We use an LCoS-SLM to apply a static phase mask that compensates for the residual phase error within the array (see Methods).}

Upon focusing, the 2D phased array generates a diffraction-limited spot whose position within the first Brillouin zone is controlled by $(\varphi_x,\varphi_y)$, which \emph{is} dependent on laser frequency $\nu$.
We leverage this \review{frequency-to-position} mapping to synthesize arbitrary spatial patterns with the cascaded RIPA-SLM. \review{In fact, this frequency-domain encoding suppresses low-frequency interference that would otherwise compromise the fidelity of the target pattern (see Methods).}

When the single-frequency phased array from the cascaded RIPA device is focused and imaged, the measured interference pattern exhibits a single, diffraction-limited spot (Fig.~\ref{fig:fig3}\textbf{b}), in analogy to the 1D case.
Sweeping the laser frequency $\nu$ rasters this spot across the image plane row-by-row (depicted schematically in Fig.~\ref{fig:fig3}\textbf{a}).
A properly chosen set of optical tones produces the ``S''-shaped pattern observed in Fig.~\ref{fig:fig3}\textbf{c}. It is composed of 11 individual spots, each corresponding to a distinct optical frequency $\nu_k$. 
These frequency components are synthesized by phase-modulating a stabilized carrier ($\nu^\mathrm{carrier}$) using a broadband EOM which is itself driven by an arbitrary waveform generator (AWG). The AWG generates a multi-tone RF field $E(t)=\sum_k A_k\cos(2\pi\nu_k^\mathrm{mod}t+\phi_k)$, consisting of a sum of frequency tones $\{\nu_k^\mathrm{mod}\}$ with amplitudes $\{A_k\}$ and phases $\{\phi_k\}$, thereby generating optical sidebands at $\nu_k=\nu^\mathrm{carrier}+ n\cdot\nu_k^\mathrm{mod}\,(n\in \mathbb{Z})$ (Fig.~\ref{fig:fig3}\textbf{d}). 
We then employ a flat-top, sharp roll-off $2^{\mathrm{nd}}$-order optical filter~\cite{Li2025Filter} ($\sim$~4~GHz bandwidth, see Methods) to suppress the optical carrier and unwanted modulation orders, selecting only the $n=+1$ order for injection into the RIPA. 

In Fig.~\ref{fig:fig3}\textbf{e}, we characterize the system crosstalk \review{in the image plane where the spots are formed}, by measuring the weighted optical power located at a \review{transverse} distance $d$ from a generated spot \review{(see Methods)}. Under the choice of different imaging systems, the \review{transverse} distance $d$ simultaneously scales with the spot waist $w_0^\prime$ ($1/e^2$ intensity radius), thus the crosstalk naturally depends on the normalized \review{transverse} distance $d/w_0^\prime$. 
Due to the \review{discreteness of the phased array}, the crosstalk exhibits power-law tails measured to be $c(d)\sim \left(\frac{d}{w_0^\prime}\right)^{-3.1}$ at large distances (see Supplementary Information for details), rather than the exponential decay \review{$c(d)\sim \exp\left(-\frac{2d^2}{{w_0^\prime}^2}\right)$} of a Gaussian beam. \review{This power-law tail is comparable to the $c(d)\sim d^{-2}$ tail expected from the Fourier transform of a clipped Gaussian beam in crossed-AOD systems~\cite{decruyenaere2026fast,goodman1969introduction}.}
The azimuthal average of this crosstalk reaches as low as $2.8\times 10^{-3}$ at a separation of $d/w_0^\prime=6.1$.
These results agree well with our theoretical model, with which we project a crosstalk of $10^{-3}$ at a separation of $6.5$ waists ($d/w_0^\prime=6.5$) and $10^{-4}$ at a separation of $13.5$ waists for a future \review{$80\times80$} RIPA modulator (see Methods).

The system also exhibits high uniformity, with the interference spot shape remaining largely invariant across the tuning range. This is expected given that the operating frequency range $\mathrm{FSR}_1\approx3$~GHz is negligible compared \review{with the frequency scale of} material dispersion, and the fact that the phased array maintains a fixed optical path independent of the laser frequency $\nu$.
Uniformity maps (Fig.~\ref{fig:fig3}\textbf{f}-\textbf{h}) confirm a peak intensity standard deviation of $2.6\%$ that could be further suppressed by finer control of RF signal amplitudes, while beam waist variations ($3.1\%$ in $x$, $1.9\%$ in $y$) remain below the camera pixel size.

\section*{Dynamic Operation and Characterization of the RIPA-SLM}
As illustrated in Fig.~\ref{fig:fig4}\textbf{a}, we can move beyond the \emph{static} operation of the RIPA-SLM by encoding arbitrary amplitude- or phase- \emph{modulation} onto the existing frequency tones, which the RIPA-SLM directly maps onto amplitude- or phase- modulation of the corresponding spots.
Such dynamic operations leverage frequency-bin encoding, which is intrinsically crosstalk-free (up to the Fourier limit) and supports parallel, independent control.
The RIPA-SLM thus promises a modulation bandwidth \review{up to the frequency-channel spacing}, fundamentally governed by the spectral resolution of the constituent spectrometer, specifically $f_{\mathrm{res}}=c/(N_x L_{\mathrm{rt,2}})=16~\mathrm{MHz}$.
Beyond discrete addressing, continuous frequency \textit{sweeping} facilitates smooth motion of the corresponding spots along the fast-moving axis in the raster scan pattern.

In Fig.~\ref{fig:fig4}, we characterize the temporal response of the RIPA-SLM and demonstrate its continuous operation with dynamical reconfiguration. Since the RIPA-SLM operates faster than any commercially available camera~\cite{li2025mega}, we employ a high-speed photodetector (PD) on a 2D translation stage as our ``camera'' (see Methods) to resolve the optical dynamics. This approach achieves a 2~ns temporal resolution and a $5~\mu$m spatial resolution.
We first demonstrate random-access addressing across a four-site registry arranged in a line, as shown in Fig.~\ref{fig:fig4}\textbf{b}. The plot shows stacked line-cuts through the registry with 4~ns spacing, demonstrating fast and independent local control with a 200~ns switching time and negligible cross-site interference. 

We further examine the switching speed of the RIPA-SLM by pulsing a single frequency component.
As shown in Fig.~\ref{fig:fig4}\textbf{c}, we drive the EOM with a 1~$\mu$s pulse and record the PD response integrated over a $(75~\mu\mathrm{m})^2$ region (central insets) and observe a smooth and rapid transition at the pulse edge. The 10\%-90\% rise- and fall- times are measured to be 44(1)~ns, showing shot-to-shot stability across all spatial locations. 
During the ``on'' state, the intensity settles to within 0.68\% of its peak value.
The measured 44~ns rise-time and the associated temporal dynamics are in quantitative agreement with our theoretical model, which is dominated by the 7.7~ns round-trip propagation time in the second RIPA and the 50~MHz photodetector bandwidth (see Methods). 
Furthermore, the beam maintains a well-defined spatial profile during both the ``on'' state and the transient period (central insets of Fig.~\ref{fig:fig4}\textbf{c}).

Finally, we showcase complex spatio-temporal modulation by independently moving two spots across the field of view.
As shown in Fig.~\ref{fig:fig4}\textbf{d}, linear RF frequency ramps drive continuous motion of two spots across the image plane over 200~ns, without shape distortion or mutual interference.
This confirms its capability of parallel, continuous movement of beams with arbitrary non-outer-product patterns. 
The lensing effect~\cite{dickson1972optical} arising from frequency chirping is calculated and numerically confirmed to be two orders of magnitude smaller than in AODs (see Methods).
In Fig.~\ref{fig:fig4}\textbf{e}, we split a single spot into two spots and then recombine them to examine a scenario where spots intersect. Stacked line-cuts along the beam splitting direction with 4~ns spacing reveal no substantial interference, 
\review{as the beat frequency rapidly increases beyond the detection bandwidth and is further suppressed by reduced spatial overlap at large separations.}
Moreover, any residual intensity oscillations should occur at the beat frequency (here 78~MHz).
For atomic physics applications, these oscillations are fast compared to sub-MHz atom motion and MHz gates, and thus average out.

\section*{Outlook}
\label{sec:outlook}
We demonstrate a high-speed two-dimensional optical display capable of performing fully-independent site-resolved addressing at an effective frame rate exceeding 10~MHz. The device is powered by a first-of-its-kind ultra-high-resolution spectrometer based on a novel interferometric architecture---the \ripa, which achieves its spectral sensitivity through long path lengths and stabilizes its spatial mode though re-imaging lens-guides.
\review{The device implements high-bandwidth single-wire electrical modulation (encoding) in the RF domain and low-crosstalk frequency-to-position mapping (decoding) in the optical domain.}

By minimizing round-trip power loss and adopting custom micro-fabricated optics, scaling the current system to an \review{$80\times80$} array and beyond \review{can be achieved without increasing the total footprint (see Methods). Specifically, at fixed per-site modulation bandwidth, the optical path length that dominates the footprint scales \emph{down} approximately as $1/N_{x,y}$.} \review{Megapixel-scale resolution can be achieved by combining multiple individually modulated wavelengths, each addressing a distinct free-spectral range of the RIPA (see Supplementary Information)}.
\review{Such a high-resolution device} unlocks numerous application domains:
Foremost among these is the realization of optical tweezer arrays with dynamically reconfigurable trapping potentials \review{and arbitrary \emph{in-situ} gates} for neutral atom quantum computing, where the RIPA-SLM provides scalable, parallel, and independent atom \review{control} \review{(see Methods and Ext. Data Fig.~\ref{efig:fige6})} to facilitate efficient quantum error correction (QEC) schemes~\cite{guo2025towards,bravyi2024high,hong2024long}, logic qubit operation~\cite{Chen2026Transversal,horsman2012surface}, magic state cultivation~\cite{gidney2024magic,sahay2025fold}, 
and continuous atom replacement~\cite{Chow2024,chiu2025continuous,li2025fastcontinuouscoherentatom}.
Furthermore, integration with recent control techniques~\cite{guo2025acoustoopticlens3dshuttling,picard2025threedimensionalacoustoopticdeflector} will enable high-speed, collision-free atom rearrangement.

Beyond quantum information science, the RIPA has myriad applications ranging from ultra-precision spectroscopy~\cite{nugent2012mid,sadiek2024airspaced,liang2025modulated} and narrow-band frequency multiplexing for quantum networks~\cite{wengerowsky2018entanglement,komza2025multiplexed,sinclair2014spectral,pu2017experimental} to
MHz-rate random-access volumetric microscopy~\cite{katona2012fast,yamaguchi2023multi}, all-optical neural interrogation for optogenetics~\cite{katona2012fast}, as well as agile beam steering for next-generation free-space communication and LiDAR~\cite{lin2022high,hail2025ultrafast}. 

\review{Note added in proof: After submission of this work, we became aware of a concurrent study on the use of a VIPA device for pattern generation~\cite{bytyqi2026devicemhzraterasteringarbitrary}.}

\section*{Acknowledgments}
We thank Hengyun Zhou, Bichen Zhang, Bowen Li for helpful discussions and feedback during the drafting of this paper.
We thank Marius Jürgensen for his valuable input on the geometric design.
This work was supported by the Office of Naval Research (ONR) through Grant N000142512291, and by the National Science Foundation (NSF) through QLCI-HQAN grant 2016136.
Z.L. acknowledges support from the Urbanek-Chodorow Fellowship. 
A.L.S. is supported by the Stanford Science Fellowship, and additionally by the Felix Bloch Fellowship and the Urbanek-Chodorow Fellowship.

\section*{Author Contributions}
X.W., Z.L., and A.V.K. performed the experiments. X.W., Z.L, A.L.S., A.V.K. contributed to data analysis. X.W, Z.L, J.S., and D.I.S. contributed to prototypes of the experimental geometry. X.W., Z.L. and J.S. wrote the manuscript with contributions and input from all authors. D.I.S. and J.S. supervised this project. X.W, and Z.L. contributed equally.

\section*{Competing Interests}
J.S. acts as a consultant to and holds stock options from Atom Computing. X.W, Z.L, A.V.K, D.I.S. and J.S. hold a patent on the phased array geometry described in this work.

\bibliographystyle{naturemag}
\bibliography{references}

@article{von1960properties,
  title={Properties of phased arrays},
  author={Von Aulock, Wilhelm H},
  journal={Proceedings of the IRE},
  volume={48},
  number={10},
  pages={1715--1727},
  year={1960},
  publisher={IEEE}
}

@misc{gidney2024magic,
      title={Magic state cultivation: growing T states as cheap as CNOT gates}, 
      author={Craig Gidney and Noah Shutty and Cody Jones},
      year={2024},
      eprint={2409.17595},
      archivePrefix={arXiv},
      primaryClass={quant-ph},
      url={https://arxiv.org/abs/2409.17595}, 
}

@article{horsman2012surface,
  title={Surface code quantum computing by lattice surgery},
  author={Horsman, Dominic and Fowler, Austin G and Devitt, Simon and Meter, Rodney Van},
  journal={New Journal of Physics},
  volume={14},
  number={12},
  pages={123011},
  year={2012},
  doi={10.1088/1367-2630/14/12/123011},
  publisher={IOP Publishing}
}

@article{Menssen2023Scalable,
author = {Adrian J. Menssen and Artur Hermans and Ian Christen and Thomas Propson and Chao Li and Andrew J. Leenheer and Matthew Zimmermann and Mark Dong and Hugo Larocque and Hamza Raniwala and Gerald Gilbert and Matt Eichenfield and Dirk R. Englund},
journal = {Optica},
number = {10},
pages = {1366--1372},
publisher = {Optica Publishing Group},
title = {Scalable photonic integrated circuits for high-fidelity light control},
volume = {10},
month = {Oct},
year = {2023},
doi = {10.1364/OPTICA.489504}
}

@article{panuski2022full,
  title={A full degree-of-freedom spatiotemporal light modulator},
  author={Panuski, Christopher L and Christen, Ian and Minkov, Momchil and Brabec, Cole J and Trajtenberg-Mills, Sivan and Griffiths, Alexander D and McKendry, Jonathan JD and Leake, Gerald L and Coleman, Daniel J and Tran, Cung and others},
  journal={Nature Photonics},
  volume={16},
  number={12},
  pages={834--842},
  year={2022},
  doi={10.1038/s41566-022-01086-9},
  publisher={Nature Publishing Group UK London}
}

@misc{tara2026complex,
      title={Complex wavefront engineering via decoupled space-time modulation}, 
      author={Virat Tara and Anna Wirth-Singh and Johannes E. Fröch and Arka Majumdar},
      year={2026},
      eprint={2605.14468},
      archivePrefix={arXiv},
      primaryClass={physics.optics},
      url={https://arxiv.org/abs/2605.14468}, 
}

@article{bluvstein2026fault,
  title={A fault-tolerant neutral-atom architecture for universal quantum computation},
  author={Bluvstein, Dolev and Geim, Alexandra A and Li, Sophie H and Evered, Simon J and Bonilla Ataides, J Pablo and Baranes, Gefen and Gu, Andi and Manovitz, Tom and Xu, Muqing and Kalinowski, Marcin and others},
  journal={Nature},
  volume={649},
  number={8095},
  pages={39--46},
  year={2026},
  publisher={Nature Publishing Group UK London}
}

@article{bakr2009quantum,
  title={A quantum gas microscope for detecting single atoms in a Hubbard-regime optical lattice},
  author={Bakr, Waseem S and Gillen, Jonathon I and Peng, Amy and F{\"o}lling, Simon and Greiner, Markus},
  journal={Nature},
  volume={462},
  number={7269},
  pages={74--77},
  year={2009},
  doi={10.1038/nature08482},
  publisher={Nature Publishing Group UK London}
}

@article{xu2025neutral,
  title={A neutral-atom Hubbard quantum simulator in the cryogenic regime},
  author={Xu, Muqing and Kendrick, Lev Haldar and Kale, Anant and Gang, Youqi and Feng, Chunhan and Zhang, Shiwei and Young, Aaron W and Lebrat, Martin and Greiner, Markus},
  journal={Nature},
  volume={642},
  number={8069},
  pages={909--915},
  year={2025},
  publisher={Nature Publishing Group UK London}
}

@article{opencv_library,
    author = {Bradski, G.},
    journal = {Dr. Dobb's Journal of Software Tools},
    title = {{The OpenCV Library}},
    year = {2000}
}

@misc{picard2025threedimensionalacoustoopticdeflector,
      title={A three-dimensional acousto-optic deflector}, 
      author={Lewis R. B. Picard and Manuel Endres},
      year={2025},
      eprint={2510.07633},
      archivePrefix={arXiv},
      primaryClass={physics.optics}
}

@misc{guo2025acoustoopticlens3dshuttling,
      title={Acousto-optic lens for 3D shuttling of atoms in a neutral atom quantum computer}, 
      author={Zhichao Guo and Rik A. H. van Herk and Edgar J. D. Vredenbregt and Servaas J. J. M. F. Kokkelmans},
      year={2025},
      eprint={2510.09398},
      archivePrefix={arXiv},
      primaryClass={physics.atom-ph}
}

@article{Yan2023,
  title = {Superradiant and Subradiant Cavity Scattering by Atom Arrays},
  author = {Yan, Zhenjie and Ho, Jacquelyn and Lu, Yue-Hui and Masson, Stuart J. and Asenjo-Garcia, Ana and Stamper-Kurn, Dan M.},
  journal = {Phys. Rev. Lett.},
  volume = {131},
  issue = {25},
  pages = {253603},
  numpages = {7},
  year = {2023},
  month = {Dec},
  publisher = {American Physical Society},
  doi = {10.1103/PhysRevLett.131.253603}
}

@article{guo2025towards,
  title = {Toward Self-Correcting Quantum Codes for Neutral Atom Arrays},
  author = {Guo, Jinkang and Hong, Yifan and Kaufman, Adam and Lucas, Andrew},
  journal = {PRX Quantum},
  volume = {7},
  issue = {1},
  pages = {010301},
  numpages = {26},
  year = {2026},
  month = {Jan},
  publisher = {American Physical Society},
  doi = {10.1103/mfmt-fwkg}
}

@article{pogorelov2025experimental,
  title={Experimental fault-tolerant code switching},
  author={Pogorelov, Ivan and Butt, Friederike and Postler, Lukas and Marciniak, Christian D and Schindler, Philipp and M{\"u}ller, Markus and Monz, Thomas},
  journal={Nature Physics},
  volume={21},
  number={2},
  pages={298--303},
  year={2025},
  publisher={Nature Publishing Group UK London}
}

@article{bluvstein2022quantum,
  title={A quantum processor based on coherent transport of entangled atom arrays},
  author={Bluvstein, Dolev and Levine, Harry and Semeghini, Giulia and Wang, Tout T and Ebadi, Sepehr and Kalinowski, Marcin and Keesling, Alexander and Maskara, Nishad and Pichler, Hannes and Greiner, Markus and others},
  journal={Nature},
  volume={604},
  number={7906},
  pages={451--456},
  year={2022},
  publisher={Nature Publishing Group UK London}
}

@article{yamaguchi2023multi,
  title={Multi-neuronal recording in unrestrained animals with all acousto-optic random-access line-scanning two-photon microscopy},
  author={Yamaguchi, Akihiro and Wu, Rui and McNulty, Paul and Karagyozov, Doycho and Mihovilovic Skanata, Mirna and Gershow, Marc},
  journal={Frontiers in Neuroscience},
  volume={17},
  pages={1135457},
  year={2023},
  doi={10.3389/fnins.2023.1135457},
  publisher={Frontiers Media SA}
}

@article{wengerowsky2018entanglement,
  title={An entanglement-based wavelength-multiplexed quantum communication network},
  author={Wengerowsky, S{\"o}ren and Joshi, Siddarth Koduru and Steinlechner, Fabian and H{\"u}bel, Hannes and Ursin, Rupert},
  journal={Nature},
  volume={564},
  number={7735},
  pages={225--228},
  year={2018},
  publisher={Nature Publishing Group UK London}
}

@article{sinclair2014spectral,
  title = {Spectral Multiplexing for Scalable Quantum Photonics using an Atomic Frequency Comb Quantum Memory and Feed-Forward Control},
  author = {Sinclair, Neil and Saglamyurek, Erhan and Mallahzadeh, Hassan and Slater, Joshua A. and George, Mathew and Ricken, Raimund and Hedges, Morgan P. and Oblak, Daniel and Simon, Christoph and Sohler, Wolfgang and Tittel, Wolfgang},
  journal = {Phys. Rev. Lett.},
  volume = {113},
  issue = {5},
  pages = {053603},
  numpages = {5},
  year = {2014},
  month = {Jul},
  publisher = {American Physical Society},
  doi = {10.1103/PhysRevLett.113.053603}
}

@article{komza2025multiplexed,
author = {Lukasz Komza and Xueyue Zhang and Hanbin Song and Yu-Lung Tang and Xin Wei and Alp Sipahigil},
journal = {Optica},
number = {9},
pages = {1400--1405},
publisher = {Optica Publishing Group},
title = {Multiplexed color centers in a silicon photonic cavity array},
volume = {12},
month = {Sep},
year = {2025},
doi = {10.1364/OPTICA.564691},
}

@article{eastin2009restrictions,
  title = {Restrictions on Transversal Encoded Quantum Gate Sets},
  author = {Eastin, Bryan and Knill, Emanuel},
  journal = {Phys. Rev. Lett.},
  volume = {102},
  issue = {11},
  pages = {110502},
  numpages = {4},
  year = {2009},
  month = {Mar},
  publisher = {American Physical Society},
  doi = {10.1103/PhysRevLett.102.110502}
}

@article{Orsi2024,
  title = {Cavity Microscope for Micrometer-Scale Control of Atom-Photon Interactions},
  author = {Orsi, Francesca and Sauerwein, Nick and Bhatt, Rohit Prasad and Faltinath, Jonas and Fedotova, Ekaterina and Reiter, Nicola and Cantat-Moltrecht, Tigrane and Brantut, Jean-Philippe},
  journal = {PRX Quantum},
  volume = {5},
  issue = {4},
  pages = {040333},
  numpages = {13},
  year = {2024},
  month = {Dec},
  publisher = {American Physical Society},
  doi = {10.1103/PRXQuantum.5.040333}
}

@article{baum2022optical,
  title={Optical mode conversion via spatiotemporally modulated atomic susceptibility},
  author={Baum, Claire and Jaffe, Matt and Palm, Lukas and Kumar, Aishwarya and Simon, Jonathan},
  journal={Optics Express},
  volume={31},
  number={1},
  pages={528--535},
  year={2022},
  doi = {10.1364/OE.476638},
  publisher={Optica Publishing Group}
}

@article{marsh2025multimode,
  title = {Multimode Cavity QED Ising Spin Glass},
  author = {Marsh, Brendan P. and Schuller, David Atri and Ji, Yunpeng and Hunt, Henry S. and Socolof, Giulia Z. and Bowman, Deven P. and Keeling, Jonathan and Lev, Benjamin L.},
  journal = {Phys. Rev. Lett.},
  volume = {135},
  issue = {16},
  pages = {160403},
  numpages = {7},
  year = {2025},
  month = {Oct},
  publisher = {American Physical Society},
  doi = {10.1103/x19r-pzyb}
}

@article{lam2021demonstration,
  title = {Demonstration of Quantum Brachistochrones between Distant States of an Atom},
  author = {Lam, Manolo R. and Peter, Natalie and Groh, Thorsten and Alt, Wolfgang and Robens, Carsten and Meschede, Dieter and Negretti, Antonio and Montangero, Simone and Calarco, Tommaso and Alberti, Andrea},
  journal = {Phys. Rev. X},
  volume = {11},
  issue = {1},
  pages = {011035},
  numpages = {17},
  year = {2021},
  month = {Feb},
  publisher = {American Physical Society},
  doi = {10.1103/PhysRevX.11.011035}
}

@article{hong2024long,
  title = {Long-range-enhanced surface codes},
  author = {Hong, Yifan and Marinelli, Matteo and Kaufman, Adam M. and Lucas, Andrew},
  journal = {Phys. Rev. A},
  volume = {110},
  issue = {2},
  pages = {022607},
  numpages = {21},
  year = {2024},
  month = {Aug},
  publisher = {American Physical Society},
  doi = {10.1103/PhysRevA.110.022607}
}

@article{Chow2024,
  title = {Circuit-Based Leakage-to-Erasure Conversion in a Neutral-Atom Quantum Processor},
  author = {Chow, Matthew N. H. and Buchemmavari, Vikas and Omanakuttan, Sivaprasad and Little, Bethany J. and Pandey, Saurabh and Deutsch, Ivan H. and Jau, Yuan-Yu},
  journal = {PRX Quantum},
  volume = {5},
  issue = {4},
  pages = {040343},
  numpages = {14},
  year = {2024},
  month = {Dec},
  publisher = {American Physical Society},
  doi = {10.1103/PRXQuantum.5.040343}
}

@article{radnaev2025universal,
  title = {Universal Neutral-Atom Quantum Computer with Individual Optical Addressing and Nondestructive Readout},
  author = {Radnaev, A.G. and Chung, W.C. and Cole, D.C. and Mason, D. and Ballance, T.G. and Bedalov, M.J. and Belknap, D.A. and Berman, M.R. and Blakely, M. and Bloomfield, I.L. and Buttler, P.D. and Campbell, C. and Chopinaud, A. and Copenhaver, E. and Dawes, M.K. and Eubanks, S.Y. and Friss, A.J. and Garcia, D.M. and Gilbert, J. and Gillette, M. and Goiporia, P. and Gokhale, P. and Goldwin, J. and Goodwin, D. and Graham, T.M. and Guttormsson, C.J. and Hickman, G.T. and Hurtley, L. and Iliev, M. and Jones, E.B. and Jones, R.A. and Kuper, K.W. and Lewis, T.B. and Lichtman, M.T. and Majdeteimouri, F. and Mason, J.J. and McMaster, J.K. and Miles, J.A. and Mitchell, P.T. and Murphree, J.D. and Neff-Mallon, N.A. and Oh, T. and Omole, V. and Parlo Simon, C. and Pederson, N. and Perlin, M.A. and Reiter, A. and Rines, R. and Romlow, P. and Scott, A.M. and Stiefvater, D. and Tanner, J.R. and Tucker, A.K. and Vinogradov, I.V. and Warter, M.L. and Yeo, M. and Saffman, M. and Noel, T.W.},
  journal = {PRX Quantum},
  volume = {6},
  issue = {3},
  pages = {030334},
  numpages = {20},
  year = {2025},
  month = {Aug},
  publisher = {American Physical Society},
  doi = {10.1103/66s8-jj18}
}

@article{graham2022multi,
  title={Multi-qubit entanglement and algorithms on a neutral-atom quantum computer},
  author={Graham, TM and Song, Y and Scott, J and Poole, C and Phuttitarn, L and Jooya, K and Eichler, P and Jiang, X and Marra, A and Grinkemeyer, B and others},
  journal={Nature},
  volume={604},
  number={7906},
  pages={457--462},
  year={2022},
  doi={10.1038/s41586-022-04603-6},
  publisher={Nature Publishing Group UK London}
}

@article{bravyi2024high,
  title={High-threshold and low-overhead fault-tolerant quantum memory},
  author={Bravyi, Sergey and Cross, Andrew W and Gambetta, Jay M and Maslov, Dmitri and Rall, Patrick and Yoder, Theodore J},
  journal={Nature},
  volume={627},
  number={8005},
  pages={778--782},
  year={2024},
  doi={10.1038/s41586-024-07107-7},
  publisher={Nature Publishing Group UK London}
}

@article{chiu2025continuous,
  title={Continuous operation of a coherent 3,000-qubit system},
  author={Chiu, Neng-Chun and Trapp, Elias C and Guo, Jinen and Abobeih, Mohamed H and Stewart, Luke M and Hollerith, Simon and Stroganov, Pavel L and Kalinowski, Marcin and Geim, Alexandra A and Evered, Simon J and others},
  journal={Nature},
  volume={646},
  pages={1075-–1080},
  year={2025},
  doi={10.1038/s41586-025-09596-6},
  publisher={Nature Publishing Group UK London}
}

@article{katona2012fast,
  title={Fast two-photon in vivo imaging with three-dimensional random-access scanning in large tissue volumes},
  author={Katona, Gergely and Szalay, Gergely and Ma{\'a}k, P{\'a}l and Kasz{\'a}s, Attila and Veress, M{\'a}t{\'e} and Hillier, D{\'a}niel and Chiovini, Bal{\'a}zs and Vizi, E Sylvester and Roska, Botond and R{\'o}zsa, Bal{\'a}zs},
  journal={Nature methods},
  volume={9},
  number={2},
  pages={201--208},
  year={2012},
  doi={10.1038/nmeth.1851},
  publisher={Nature Publishing Group US New York}
}

@misc{hail2025ultrafast,
      title={Ultrafast, reconfigurable all-optical beam steering and spatial light modulation}, 
      author={Claudio U. Hail and Lior Michaeli and Harry A. Atwater},
      year={2025},
      eprint={2511.03860},
      archivePrefix={arXiv},
      primaryClass={physics.optics}
}

@article{lin2022high,
  title={High-performance optical beam steering with nanophotonics},
  author={Lin, Sam and Chen, Yixin and Wong, Zi Jing},
  journal={Nanophotonics},
  volume={11},
  number={11},
  pages={2617--2638},
  year={2022},
  doi={10.1515/nanoph-2021-0805/html},
  publisher={De Gruyter}
}

@article{bluvstein2024logical,
  title={Logical quantum processor based on reconfigurable atom arrays},
  author={Bluvstein, Dolev and Evered, Simon J and Geim, Alexandra A and Li, Sophie H and Zhou, Hengyun and Manovitz, Tom and Ebadi, Sepehr and Cain, Madelyn and Kalinowski, Marcin and Hangleiter, Dominik and others},
  journal={Nature},
  volume={626},
  number={7997},
  pages={58--65},
  year={2024},
  doi={10.1038/s41586-023-06927-3},
  publisher={Nature Publishing Group UK London}
}

@article{Chiu2018,
  title = {Quantum State Engineering of a Hubbard System with Ultracold Fermions},
  author = {Chiu, Christie S. and Ji, Geoffrey and Mazurenko, Anton and Greif, Daniel and Greiner, Markus},
  journal = {Phys. Rev. Lett.},
  volume = {120},
  issue = {24},
  pages = {243201},
  numpages = {6},
  year = {2018},
  month = {Jun},
  publisher = {American Physical Society},
  doi = {10.1103/PhysRevLett.120.243201}
}

@article{steinhauer2016observation,
  title={Observation of quantum Hawking radiation and its entanglement in an analogue black hole},
  author={Steinhauer, Jeff},
  journal={Nature Physics},
  volume={12},
  number={10},
  pages={959--965},
  year={2016},
  doi={10.1038/nphys3863},
  publisher={Nature Publishing Group UK London}
}

@article{pu2017experimental,
  title={Experimental realization of a multiplexed quantum memory with 225 individually accessible memory cells},
  author={Pu, YF and Jiang, Nan and Chang, Wei and Yang, HX and Li, Chang and Duan, LM},
  journal={Nature communications},
  volume={8},
  number={1},
  pages={15359},
  year={2017},
  doi={10.1038/ncomms15359},
  publisher={Nature Publishing Group UK London}
}

@article{clark2017collective,
  title={Collective emission of matter-wave jets from driven Bose--Einstein condensates},
  author={Clark, Logan W and Gaj, Anita and Feng, Lei and Chin, Cheng},
  journal={Nature},
  volume={551},
  number={7680},
  pages={356--359},
  year={2017},
  doi={10.1038/nature24272},
  publisher={Nature Publishing Group UK London}
}

@article{Yamazaki2010,
  title = {Submicron Spatial Modulation of an Interatomic Interaction in a Bose-Einstein Condensate},
  author = {Yamazaki, Rekishu and Taie, Shintaro and Sugawa, Seiji and Takahashi, Yoshiro},
  journal = {Phys. Rev. Lett.},
  volume = {105},
  issue = {5},
  pages = {050405},
  numpages = {4},
  year = {2010},
  month = {Jul},
  publisher = {American Physical Society},
  doi = {10.1103/PhysRevLett.105.050405}
}

@article{takeda1982fourier,
  title={Fourier-transform method of fringe-pattern analysis for computer-based topography and interferometry},
  author={Takeda, Mitsuo and Ina, Hideki and Kobayashi, Seiji},
  journal={Journal of the optical society of America},
  volume={72},
  number={1},
  pages={156--160},
  year={1982},
  doi={10.1364/JOSA.72.000156},
  publisher={Optical Society of America}
}

@article{bozinovic2013terabit,
  title={Terabit-scale orbital angular momentum mode division multiplexing in fibers},
  author={Bozinovic, Nenad and Yue, Yang and Ren, Yongxiong and Tur, Moshe and Kristensen, Poul and Huang, Hao and Willner, Alan E and Ramachandran, Siddharth},
  journal={science},
  volume={340},
  number={6140},
  pages={1545--1548},
  year={2013},
  doi={10.1126/science.1237861},
  publisher={American Association for the Advancement of Science}
}

@article{richardson2013space,
  title={Space-division multiplexing in optical fibres},
  author={Richardson, David J and Fini, John M and Nelson, Lynn E},
  journal={Nature photonics},
  volume={7},
  number={5},
  pages={354--362},
  year={2013},
  publisher={Nature Publishing Group}
}

@article{fahrbach2010microscopy,
  title={Microscopy with self-reconstructing beams},
  author={Fahrbach, Florian O and Simon, Philipp and Rohrbach, Alexander},
  journal={Nature photonics},
  volume={4},
  number={11},
  pages={780--785},
  year={2010},
  doi={10.1038/s41566-021-00780-4},
  publisher={Nature Publishing Group UK London}
}

@article{Hell2007,
  title={Far-field optical nanoscopy},
  author={Hell, Stefan W.},
  journal={Science},
  volume={316},
  number={5828},
  pages={1153--1158},
  year={2007},
  publisher={American Association for the Advancement of Science}
}

@article{Vettenburg2014,
  title={Light-sheet microscopy using an Airy beam},
  author={Vettenburg, Tom and Dalgarno, H. I. C. and Nylk, J. and Coll-Llad{\'o}, C. and Ferrier, D. E. K. and {\v{C}}i{\v{z}}m{\'a}r, T. and Gunn-Moore, F. J. and Dholakia, K.},
  journal={Nature Methods},
  volume={11},
  number={5},
  pages={541--544},
  year={2014},
  doi={10.1038/nmeth.2922},
  publisher={Nature Publishing Group}
}

@article{brown2019gray,
  title={Gray-molasses optical-tweezer loading: controlling collisions for scaling atom-array assembly},
  author={Brown, MO and Thiele, T and Kiehl, C and Hsu, T-W and Regal, CA},
  journal={Physical Review X},
  volume={9},
  number={1},
  pages={011057},
  year={2019},
  doi={10.1103/PhysRevX.9.011057},
  publisher={APS}
}

@article{endres2016atom,
  title={Atom-by-atom assembly of defect-free one-dimensional cold atom arrays},
  author={Endres, Manuel and Bernien, Hannes and Keesling, Alexander and Levine, Harry and Anschuetz, Eric R and Krajenbrink, Alexandre and Senko, Crystal and Vuletic, Vladan and Greiner, Markus and Lukin, Mikhail D},
  journal={Science},
  volume={354},
  number={6315},
  pages={1024--1027},
  year={2016},
  doi={10.1126/science.aah3752},
  publisher={American Association for the Advancement of Science}
}

@article{liang2025modulated,
  title={Modulated ringdown comb interferometry for sensing of highly complex gases},
  author={Liang, Qizhong and Bisht, Apoorva and Scheck, Andrew and Schunemann, Peter G and Ye, Jun},
  journal={Nature},
  pages={941--948},
  volume={638},
  year={2025},
  doi={10.1038/s41586-024-08534-2},
  publisher={Nature Publishing Group UK London}
}

@article{nugent2012mid,
  title={Mid-infrared virtually imaged phased array spectrometer for rapid and broadband trace gas detection},
  author={Nugent-Glandorf, Lora and Neely, Tyler and Adler, Florian and Fleisher, Adam J and Cossel, Kevin C and Bjork, Bryce and Dinneen, Tim and Ye, Jun and Diddams, Scott A},
  journal={Optics letters},
  volume={37},
  number={15},
  pages={3285--3287},
  year={2012},
  doi={10.1364/OL.37.003285},
  publisher={Optical Society of America}
}

@article{sommer2016engineering,
  title={Engineering photonic floquet hamiltonians through fabry--p{\'e}rot resonators},
  author={Sommer, Ariel and Simon, Jonathan},
  journal={New Journal of Physics},
  volume={18},
  number={3},
  pages={035008},
  year={2016},
  doi={10.1088/1367-2630/18/3/035008},
  publisher={IOP Publishing}
}

@article{li2025mega,
  title={Mega-FPS low light camera},
  author={Li, Bowen and Palm, Lukas and J{\"u}rgensen, Marius and Feng, Yiming Cady and Greiner, Markus and Simon, Jon},
  journal={Optics Express},
  volume={33},
  number={15},
  pages={31096--31106},
  year={2025},
  doi={10.1364/OE.563793},
  publisher={Optica Publishing Group}
}

@article{manetsch2025tweezer,
  title={A tweezer array with 6100 highly coherent atomic qubits},
  author={Manetsch, Hannah J and Nomura, Gyohei and Bataille, Elie and Lv, Xudong and Leung, Kon H and Endres, Manuel},
  journal={Nature},
  volume={647},
  pages={60–-67},
  year={2025},
  doi={10.1038/s41586-025-09641-4},
  publisher={Nature Publishing Group UK London}
}

@article{dickson1972optical,
  title={Optical considerations for an acoustooptic deflector},
  author={Dickson, LeRoy D},
  journal={Applied Optics},
  volume={11},
  number={10},
  pages={2196--2202},
  year={1972},
  doi={10.1364/AO.11.002196},
  publisher={OSA}
}

@article{shirasaki_VIPA,
  author  = {Shirasaki, Masataka},
  title   = {Large angular dispersion by a virtually imaged phased array and its application to a wavelength demultiplexer},
  journal = {Optics Letters},
  year    = {1996},
  volume  = {21},
  number  = {5},
  pages   = {366--368},
  doi     = {10.1364/OL.21.000366}
}

@article{zupancic2016ultra,
  title={Ultra-precise holographic beam shaping for microscopic quantum control},
  author={Zupancic, Philip and Preiss, Philipp M and Ma, Ruichao and Lukin, Alexander and Eric Tai, M and Rispoli, Matthew and Islam, Rajibul and Greiner, Markus},
  journal={Optics express},
  volume={24},
  number={13},
  pages={13881--13893},
  year={2016},
  doi={10.1364/OE.24.013881},
  publisher={Optical Society of America}
}

@misc{li2025fastcontinuouscoherentatom,
      title={Fast, continuous and coherent atom replacement in a neutral atom qubit array}, 
      author={Yiyi Li and Yicheng Bao and Michael Peper and Chenyuan Li and Jeff D. Thompson},
      year={2025},
      eprint={2506.15633},
      archivePrefix={arXiv},
      primaryClass={quant-ph}
}

@misc{zhang2025leveragingerasureerrorslogical,
      title={Leveraging erasure errors in logical qubits with metastable $^{171}$Yb atoms}, 
      author={Bichen Zhang and Genyue Liu and Guillaume Bornet and Sebastian P. Horvath and Pai Peng and Shuo Ma and Shilin Huang and Shruti Puri and Jeff D. Thompson},
      year={2025},
      eprint={2506.13724},
      archivePrefix={arXiv},
      primaryClass={quant-ph}
}

@article{chan20082,
  title={2-Dimensional beamsteering using dispersive deflectors and wavelength tuning},
  author={Chan, Trevor and Myslivets, Evgeny and Ford, Joseph E},
  journal={Optics Express},
  volume={16},
  number={19},
  pages={14617--14628},
  year={2008},
  publisher={Optical Society of America}
}

@article{zhang2024scaled,
  title={Scaled local gate controller for optically addressed qubits},
  author={Zhang, Bichen and Peng, Pai and Paul, Aditya and Thompson, Jeff D},
  journal={Optica},
  volume={11},
  number={2},
  pages={227--233},
  year={2024},
  doi={10.1364/OPTICA.512155},
  publisher={Optica Publishing Group}
}

@article{sadiek2024airspaced,
  title={Air-spaced virtually imaged phased array with 94 MHz resolution for precision spectroscopy},
  author={Sadiek, Ibrahim and Lang, Norbert and Van Helden, Jean-Pierre H},
  journal={Optics Express},
  volume={32},
  number={26},
  pages={46511--46521},
  year={2024},
  doi={10.1364/OE.538725},
  publisher={Optica Publishing Group}
}

@misc{Li2025Filter,
      title={A Second-Order Optical Butterworth Fabry-P\'erot Filter}, 
      author={Zeyang Li and Abhishek V. Karve and Xin Wei and Jonathan Simon},
      year={2025},
      eprint={2510.15032},
      archivePrefix={arXiv},
      primaryClass={physics.optics}
}

@article{jaffe2022understanding,
  title={Understanding and suppressing backscatter in optical resonators},
  author={Jaffe, Matt and Palm, Lukas and Baum, Claire and Taneja, Lavanya and Kumar, Aishwarya and Simon, Jonathan},
  journal={Optica},
  volume={9},
  number={8},
  pages={878--885},
  year={2022},
  doi={10.1364/OPTICA.463723},
  publisher={Optica Publishing Group}
}

@article{shadmany2025cavity,
  title={Cavity QED in a High NA resonator},
  author={Shadmany, Danial and Kumar, Aishwarya and Soper, Anna and Palm, Lukas and Yin, Chuan and Ando, Henry and Li, Bowen and Taneja, Lavanya and Jaffe, Matt and David, Schuster and others},
  journal={Science Advances},
  volume={11},
  number={9},
  pages={eads8171},
  year={2025},
  doi={10.1126/sciadv.ads8171},
  publisher={American Association for the Advancement of Science}
}

@article{zhao2025integrated,
  title={An integrated photonics platform for high-speed, ultrahigh-extinction, many-channel quantum control},
  author={Zhao, Mengdi and Singh, Manuj and Singh, Anshuman and Thoreen, Henry and DeAngelo, Robert J and Dominguez, Daniel and Leenheer, Andrew and Peyskens, Fr{\'e}d{\'e}ric and Lukin, Alexander and Englund, Dirk and others},
  journal={arXiv preprint arXiv:2508.09920},
  year={2025}
}

@misc{Tim_optable,
author = {Wei, Xin},
license = {MIT},
title = {{optable}},
url = {https://github.com/tim4431/optable},
}

@misc{Tim_vipa_focus_simulation,
author = {Wei, Xin},
license = {MIT},
title = {{vipa focus simulation}},
url = {https://github.com/tim4431/vipa_focus_simulation},
}

@article{decruyenaere2026fast,
  title={Fast projections of two-dimensional light patterns using acousto-optical deflectors},
  author={Decruyenaere, Robbert and Tanghe, Clara and Van Wellen, Senne and Van Acoleyen, Karel},
  journal={arXiv preprint arXiv:2604.19421},
  year={2026}
}

@misc{goodman1969introduction,
  title={Introduction to Fourier optics},
  author={Goodman, Joseph W and Cox, Mary E},
  year={1969},
  publisher={American Institute of Physics}
}

@article{wang2015coherent,
  title = {Coherent Addressing of Individual Neutral Atoms in a 3D Optical Lattice},
  author = {Wang, Yang and Zhang, Xianli and Corcovilos, Theodore A. and Kumar, Aishwarya and Weiss, David S.},
  journal = {Phys. Rev. Lett.},
  volume = {115},
  issue = {4},
  pages = {043003},
  numpages = {5},
  year = {2015},
  month = {Jul},
  publisher = {American Physical Society},
  doi = {10.1103/PhysRevLett.115.043003},
}

@article{chen2026transversal,
  title={Transversal logical Clifford gates on the rotated surface code with reconfigurable neutral atom arrays},
  author={Chen, Zi-Han and Chen, Ming-Cheng and Lu, Chao-Yang and Pan, Jian-Wei},
  journal={Physical Review Letters},
  volume={136},
  number={13},
  pages={130601},
  year={2026},
  publisher={APS}
}

@article{jia2018strongly,
  title={A strongly interacting polaritonic quantum dot},
  author={Jia, Ningyuan and Schine, Nathan and Georgakopoulos, Alexandros and Ryou, Albert and Clark, Logan W and Sommer, Ariel and Simon, Jonathan},
  journal={Nature Physics},
  volume={14},
  number={6},
  pages={550--554},
  year={2018},
  publisher={Nature Publishing Group UK London}
}

@article{sahay2025fold,
  title={Fold-transversal surface code cultivation},
  author={Sahay, Kaavya and Tsai, Pei-Kai and Chang, Kathleen and Su, Qile and Smith, Thomas B and Singh, Shraddha and Puri, Shruti},
  journal={arXiv preprint arXiv:2509.05212},
  year={2025}
}

@misc{bytyqi2026devicemhzraterasteringarbitrary,
      title={Device for MHz-rate rastering of arbitrary 2D optical potentials}, 
      author={Edita Bytyqi and Josiah Sinclair and Joshua Ramette and Vladan Vuletić},
      year={2026},
      eprint={2602.16025},
      archivePrefix={arXiv},
      primaryClass={quant-ph},
      url={https://arxiv.org/abs/2602.16025}, 
}

@misc{lib2026velocity,
      title={Velocity-Enabled Quantum Computing with Neutral Atoms}, 
      author={Ohad Lib and Hendrik Timme and Maximilian Ammenwerth and Flavien Gyger and Renhao Tao and Shijia Sun and Immanuel Bloch and Johannes Zeiher},
      year={2026},
      eprint={2603.15561},
      archivePrefix={arXiv},
      primaryClass={quant-ph},
      url={https://arxiv.org/abs/2603.15561}, 
}

@misc{xue2026factoring2048bitrsa,
      title={Factoring $2048$ bit RSA integers with a half-million-qubit modular atomic processor}, 
      author={Tian Xue and Jacob P. Covey},
      year={2026},
      eprint={2605.03951},
      archivePrefix={arXiv},
      primaryClass={quant-ph},
      url={https://arxiv.org/abs/2605.03951}, 
}

@article{dudinets2025all,
  title={All-to-all connectivity of Rydberg-atom-based quantum processors with messenger qubits},
  author={Dudinets, Ivan V and Straupe, Stanislav S and Fedorov, Aleksey K and Lychkovskiy, Oleg V},
  journal={arXiv preprint arXiv:2504.05087},
  year={2025}
}





\subsection*{Data Availability}
The experimental data presented in this manuscript are available from the corresponding author upon request, due to the proprietary file formats employed in the data collection process.

\subsection*{Code Availability}
The source code for simulations and performing experiments are available from the corresponding author upon request. Codes for RIPA ray-tracing are available at Ref.~\cite{Tim_optable}. Codes for simulating the interference pattern are available at Ref.~\cite{Tim_vipa_focus_simulation}.

\clearpage
\newpage





\renewcommand{\theequation}{S\arabic{equation}}
\renewcommand{\thefigure}{\arabic{figure}}
\renewcommand{\figurename}{Ext. Data Fig.}
\renewcommand{\figurename}{Ext. Data Fig.}
\renewcommand{\tablename}{Ext. Data Table}
\setcounter{figure}{0}
\setcounter{table}{0}
\setcounter{equation}{0}

\onecolumngrid
\newpage
\phantomsection
\section*{Extended Data Figures}
\FloatBarrier

\begin{figure*}[htbp!]
	\centering
 	\includegraphics[width=183mm]{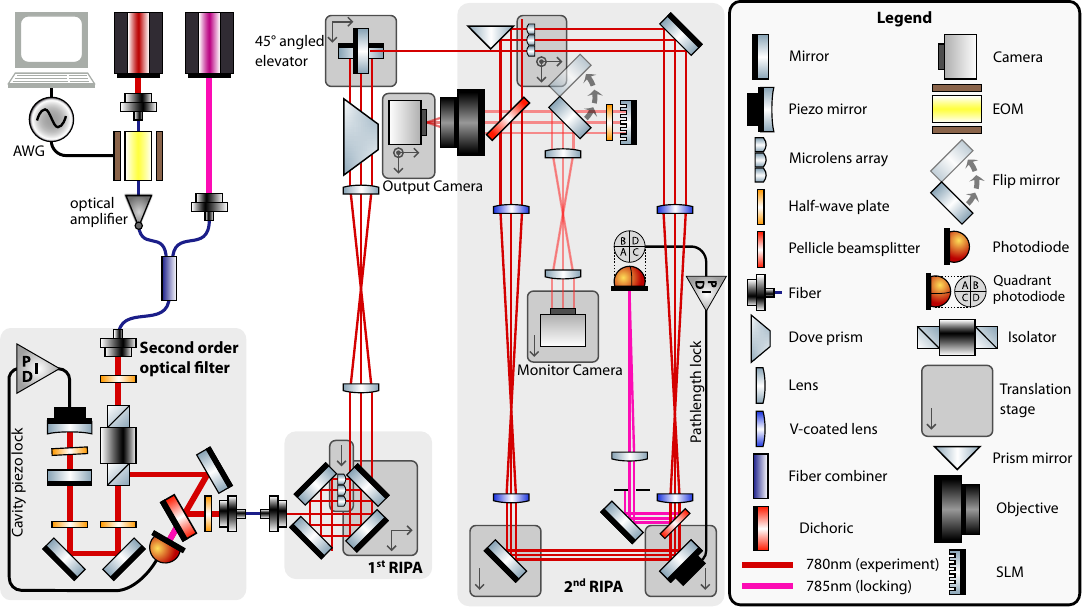}
    
	\caption{
    	\textbf{Detailed schematic of the RIPA system.}
        The experiment operates \review{at a wavelength of} 780~nm (red paths, \review{including its sidebands}), while an auxiliary \review{$\lambda=785~\mathrm{nm}$} laser (pink paths) provides \review{a reference for} active path-length stabilization. Input frequency tones are generated \review{as optical sidebands by driving an electro-optic modulator (EOM)} with an arbitrary waveform generator (AWG). \review{An optical amplifier boosts optical power after the EOM}. A second-order optical filter isolates the $+1$ modulation sidebands and injects them into the first RIPA.
        The first RIPA converts the input into a 1D beam array using a microlens array (MLA). This array is transferred via a 4f relay which contains angled elevator mirrors and a Dove prism for spatial alignment (tilt, displacement, and roll) to the second RIPA. The second RIPA employs the same MLA and V-coated 4f telescopes to generate a 2D beam array. A pellicle beamsplitter out-couples the light to a spatial light modulator (SLM) for phase calibration (see Ext. Data Fig.~\ref{efig:fige2}) and final imaging (with static phase mask).
        Active locking of the second RIPA path length is achieved via quadrant photodiodes and piezo-actuated mirrors. Pairs of mirrors mounted on translation stages allow for the independent tuning of round-trip path length \review{(for mode matching)} and beam separation \review{(for matching to the MLA pitch)}.}
	\label{efig:fige1}
\end{figure*}

\begin{figure*}[htbp!]
	\centering
 	\includegraphics[width=183mm]{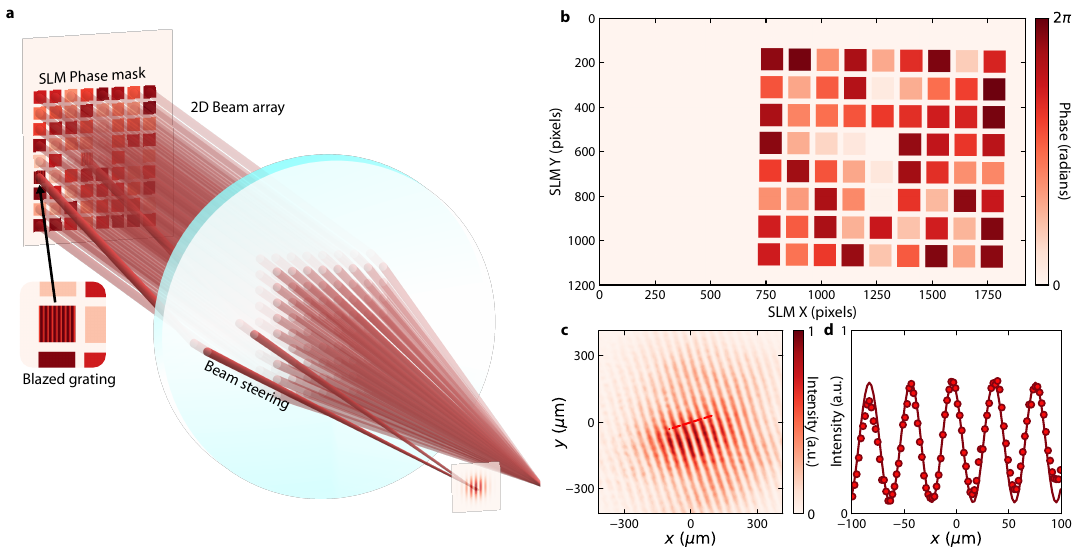}
    
	\caption{
    	\textbf{Interferometric calibration of the SLM phase mask.}
        \textbf{a}, Schematic of the phase calibration procedure. Localized blazed gratings are applied to specific sub-regions of the SLM to steer pairs of beams from the 2D beam array, causing them to interfere at the focal plane. This process compensates for \textit{static} wavefront aberrations and ensures phase uniformity across the array.
        \textbf{b}, The resulting calibrated phase mask. Each grid tile consists of $100\times100$ SLM pixels (covering an $800\times 800~\mu\mathrm{m}^2$ area), providing independent phase control for each beam in the array.
        \textbf{c}, Representative interference fringes captured at the focal plane.
        \textbf{d}, One-dimensional intensity line-cut (dots) of the fringes in \textbf{c} with a sinusoidal fit (solid line), from which the relative phase is extracted for precise compensation.}
	\label{efig:fige2}
\end{figure*}

\begin{figure*}[htbp!]
	\centering
 	\includegraphics[width=183mm]{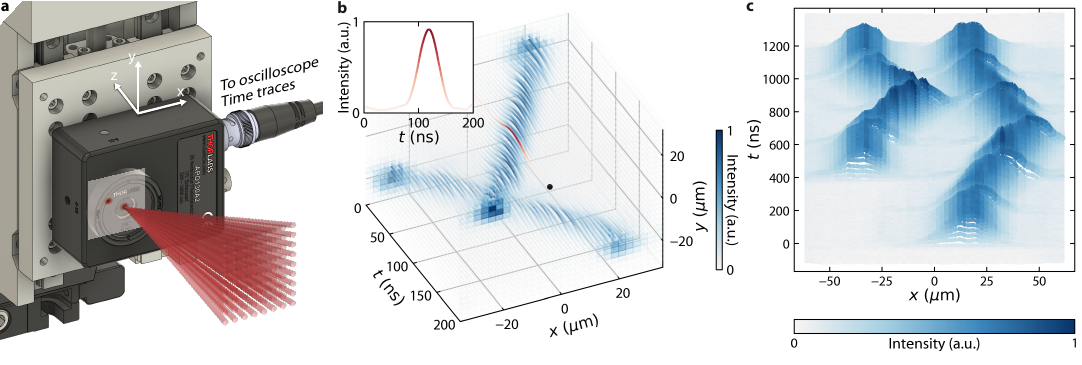}
    
	\caption{
    	\textbf{Scanning photodetector for resolving nano-second beam dynamics.}
        \textbf{a}, A photodetector (PD) with a $5~\mu\mathrm{m}$ diameter pinhole is mounted on a 3-axis translation stage to sample the focal plane. The AWG, which drives the EOM, is triggered synchronously with the oscilloscope, which records the PD voltage with 2~ns temporal resolution.
        \textbf{b}, Spatio-temporal reconstruction. Raw time-traces recorded at discrete spatial coordinates (inset) are synthesized into a spatio-temporal intensity distribution (data corresponding to Fig.~\ref{fig:fig4}\textbf{d}).
        Stacking these time traces yields a movie of spot dynamics with $5~\mu\mathrm{m} \times 5~\mu\mathrm{m}$ spatial and 2~ns temporal resolution. 
        \textbf{c}, Reconstructed 1D trajectories of spots showing the asynchronous motion of two independent spots following a zigzag trajectory.}
	\label{efig:fige3}
\end{figure*}

\begin{figure}[htbp!]
	\centering
 	\includegraphics[width=165mm]{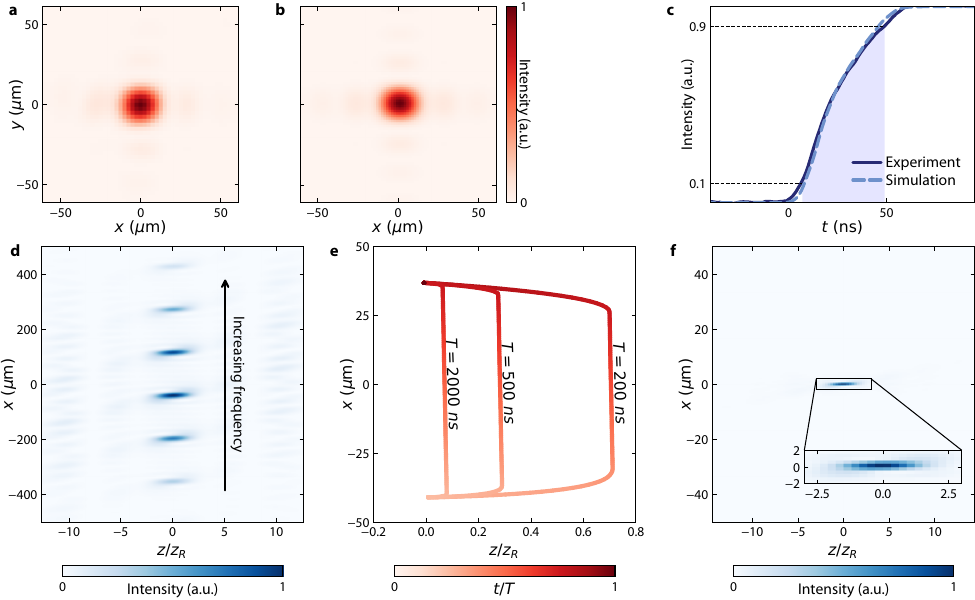}
    
	\caption{
    	\textbf{Simulated interference patterns and experimental validation.}
        \textbf{a,b}, Comparison of experimental \textbf{a} and simulated \textbf{b} focal-plane intensity distributions ($xy$ cross-section).
        \textbf{c}, Temporal evolution of the integrated intensity for the pulse sequence described in Fig.~\ref{fig:fig4}\textbf{c}. The experimental data (solid dark blue) agrees with the simulation (dashed light blue).
        \textbf{d}, Simulated beam profile in the $xz$ plane. The axial coordinate $z$ is normalized by the Rayleigh range $z_R$. The interference pattern exhibits a localized Gaussian beam profile near the focal point.
        \textbf{e}, Transient axial focal shift during linear frequency chirping. The traces show the trajectory of the beam waist in the $xz$ plane for various sweep durations ($T=$ 200, 500, and 2000~ns) for the same transverse displacement, illustrating the dynamic lensing effect.
        \textbf{f}, Simulated $xz$ cross-section of the scaled-up \review{$80\times80$} RIPA configuration. The inset provides a magnified view of the beam waist, demonstrating tight confinement.}
	\label{efig:fige4}
\end{figure}

\begin{figure*}[htbp!]
	\centering
 	\includegraphics[width=183mm]{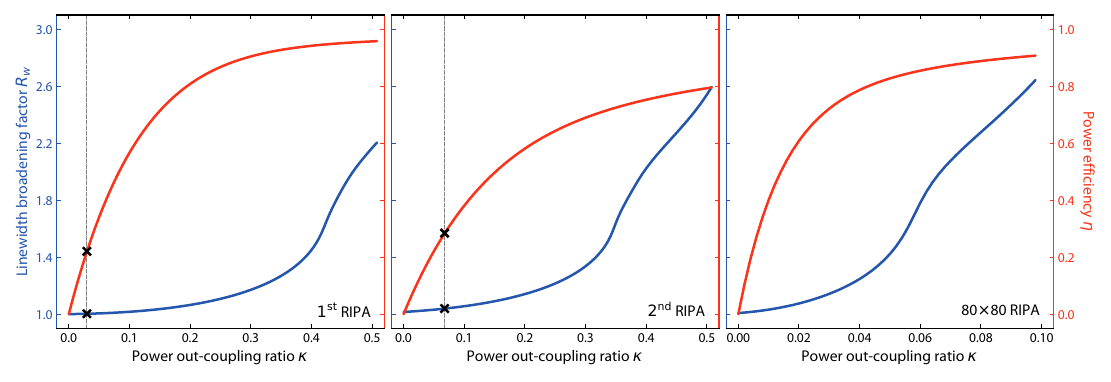}
    
	\caption{
    	\textbf{Tradeoff between power efficiency and interference linewidth.}
        Calculated \review{single-stage RIPA} power efficiency $\eta$ (red, right axis) and the interference linewidth broadening factor $R_w$ (blue, left axis) as functions of the power out-coupling ratio $\kappa$ for the first RIPA (left), the second RIPA (center), and \review{a single stage of} the proposed \review{$80\times80$} RIPA (right) configurations.
        In the first two panels, vertical dashed lines and the black crosses denote the current experimental operating points.}
	\label{efig:fige5}
\end{figure*}

\begin{figure*}[!htbp]
	\centering
 	\includegraphics[width=183mm]{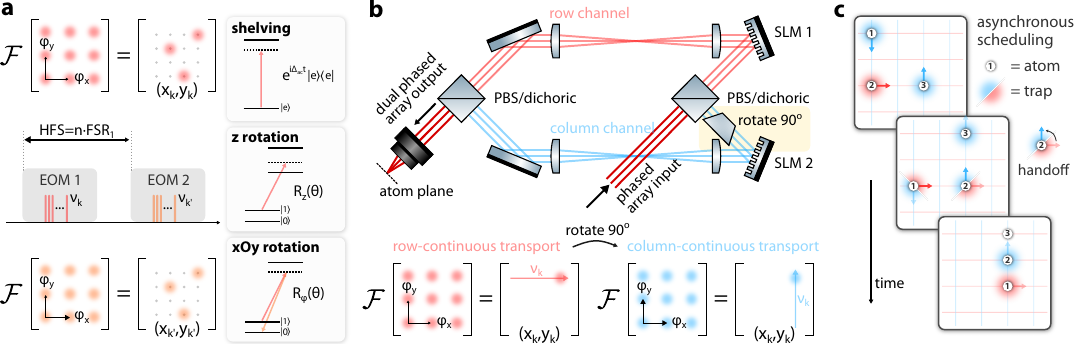}

    \caption{
    \review{
    \textbf{A RIPA-controlled quantum computer}. The RIPA-SLM enables site-resolved control and coherent transport of atomic qubits on a two-dimensional grid.
    \textbf{a}, A frequency-multiplexed phased array (red), Fourier transformed by an objective, produces local addressing beams at selected atomic sites $(x_k,y_k)$, generating site-dependent light shifts. Optical tones shifted by the hyperfine splitting (HFS), chosen to equal an integer number of RIPA free spectral ranges, address the same atoms (orange) and enable hyperfine-resolved control. The boxes on the right illustrate representative operations: scalar ac-Stark shifts for shelving, vector ac-Stark shifts for $R_z(\theta)$ rotations, and two-photon Raman transitions for $R_\varphi(\theta)$ rotations, where $\varphi$ is the quadrature angle of the rotation axis in the $xOy$ plane set by the phase difference between the two optical tones.
    \textbf{b}, Frequency ramps translate RIPA-SLM addressing spots continuously along a row, providing a mechanism for coherent atom transport. Polarization or frequency multiplexing separates the RIPA phased array output into \emph{row} and \emph{column} channels; the \emph{column} channel is rotated by $90^\circ$ relative to the \emph{row} channel, for example, using a mirror pair or Dove prism, and the two channels are then recombined and focused onto the atom plane. The resulting transport patterns move continuously along the row and column directions, respectively.
    \textbf{c}, Consecutive snapshots illustrate asynchronous transport on the grid, moving atoms from an ``L'' configuration into a ``line'' configuration. Atoms can be transported independently along the two directions and handed off at intersections of the grid lines.}
    }
	\label{efig:fige6}
\end{figure*}

\clearpage
\newpage
\twocolumngrid
\section*{Methods}
\appendix

\subsection{Description of the experiment}
Here we describe the architecture of the experimental system, focusing on frequency tone generation, the geometry of the RIPA system, and wavefront correction via a SLM. A comprehensive schematic of the layout is provided in Ext. Data Fig.~\ref{efig:fige1}.

\subsubsection*{Frequency tone generation, \review{filtering, and amplification}}
Optical frequency tones corresponding to the target spatial addressing pattern are synthesized by phase-modulating a \review{$\lambda=780~\text{nm}$} laser using an EOM (Thorlabs LNY7810A) driven by a high-speed AWG (Keysight M8195A, 65GSa/s). The AWG is programmed to generate a multi-tone RF waveform $V(t)=\frac{1}{\sum_{k=1}^N A_k}\sum_{k=1}^{N}A_k \cos\left(2\pi\nu_k^{\mathrm{mod}}\ t+\phi_k\right)$, where $\nu_k^{\mathrm{mod}}$ denotes the RF modulation frequency (centered at 11~GHz), while $A_k$ and $\phi_k$ represent the amplitude and phase of each constituent frequency tone, respectively.
This RF signal is amplified via a two-stage amplifier chain (Minicircuits ZX60-10223G+ and RF Bay MGA-22-13) to attain powers of $\approx +16\text{dBm}$ for a single tone and $\approx -7\text{dBm}$ per tone for $\sim 100$ tones.

To isolate the target spectrum, we build a second-order Butterworth-type optical filter~\cite{Li2025Filter} and actively lock it to the $+1$ sideband. 
The filter optically selects only the $+1$ modulation sidebands with a flat-passband response over a $\sim 4~\text{GHz}$ bandwidth and strongly rejects the carrier and other unwanted sidebands.

\review{Because we operate with a single-mode optical channel input, all amplitude- and frequency-domain shaping can be performed within a single spatial mode before the light is sent into the cascaded RIPAs.
We use a tapered amplifier (TA, Eagleyard, miniTA 780~nm) to boost the optical power after the modulator. The spectral gain profile of the TA is approximately flat across the few-GHz range of our optical spectrum.
The same argument also applies to potential wavelength conversion using nonlinear crystal before the cascaded RIPAs.
}

\subsubsection*{The geometry of RIPA}
As shown in Ext. Data Fig.~\ref{efig:fige1}, the experiment operates \review{at a wavelength of} 780~nm (red paths, \review{including its sidebands}), while an auxiliary \review{$\lambda=785~\text{nm}$ laser} (pink paths) \review{provides a reference} for path-length stabilization.
The first RIPA converts the input beam into a 1D beam array using a MLA that re-images the beams during each round trip. A 4f relay transfers this 1D beam array to the input of the second RIPA, where a pair of $45^\circ$ angled elevator mirrors controls the beam tilt and displacement, and a dove prism adjusts the roll angle.
The second RIPA employs an identical MLA inside its round trip, supplemented by two 4f path-length extension telescopes whose lenses are V-coated for 780~nm (reflection $< 0.1\%$) to minimize loss.
A pellicle beamsplitter (Thorlabs BP108) inside the second RIPA round trip acts as an outcoupler, directing the resulting 2D beam array at normal incidence onto a SLM for static phase compensation (as described in later sections).
The corrected phased array is then focused onto the output camera; alternatively, a flip mirror can redirect it to a monitor camera on an equivalent Fourier plane, where the beams are of equal size and propagate in parallel.

System alignment is performed by iteratively matching the beam separation to the MLA pitch ($p$) and matching the effective round-trip path length to the target round-trip length which supports the fundamental lens-guide mode. To decouple these two degrees of freedom, we utilize either mirror pairs mounted on translation stages or individual mirrors on separate motorized stages controlled via common- and differential-mode actuation.
Further details on the alignment protocol are available in Supplementary Information.

\subsubsection*{Stabilization of RIPA}
\review{As shown in Ext. Data Fig.~\ref{efig:fige1}}, the \review{$\lambda=785~\text{nm}$} auxiliary laser (pink paths) simultaneously stabilizes the optical filter cavity and the second RIPA path length. For the latter, two adjacent beams from the 2D beam array are spatially isolated and interfered at a quadrant photodiode (Thorlabs PDQ80A). The resulting error signal is fed to a piezo actuator to lock the (meter-long) round-trip path length of the second RIPA. \review{The first RIPA is passively stabilized and no substantial mechanical noise is observed.}

\review{This lock also provides control over the absolute frequency-to-position mapping. In the RIPA-SLM, relative frequency differences determine the relative positions of the addressing spots. However, the \emph{absolute} mapping is set by the RIPA round-trip phase ($\varphi_x,\varphi_y$). One can bias these phases by shifting the path-length on the order of one wavelength, experimentally achieved by changing the piezo lock point. Thus, a common shift of all optical frequencies can be compensated without moving the addressing pattern.}

\subsubsection*{SLM phase correction}
\label{sec:slm_correction}
The interference pattern of the 2D beam array is highly sensitive to optical aberrations and surface roughness, which can significantly degrade its quality.
Since the beam array is nicely discretized in space before the focusing lens, we can compensate for these phase errors immediately after the out-coupling pellicle beamsplitter using an SLM (Meadowlark E-Series, $1920\times1200$, 8-Bit) to apply a \textit{static} phase mask.
Once calibrated, this mask \review{works for all operating frequencies, and} remains stable throughout the operation of the RIPA-SLM.

The calibration procedure is performed as follows.
First, we map the 2D beam array indexed with $(i,j)$ onto the SLM pixel coordinates $(x,y)$. 
To achieve this, we employ a 4f imaging system to image the $N_x\times N_y$ array reflected from the SLM, with an iris in the Fourier plane for spatial filtering.
Applying a blazed-grating pattern to the SLM deflects specific beams so they are blocked by the iris. By narrowing the grating region via binary search and sliding-window procedures, we precisely determine the coordinate $(x,y)_{ij}$ for each beam $(i,j)$.
By determining the $(x,y)_{ij}$ for a set of selected beams (e.g., $(i,j)=(0,0),(0,N_y-1),(N_x-1,0)$), the coordinates for the entire array are obtained via an affine transformation fit.

Next, we calibrate the relative residual phase $\phi_{ij}$ between the beam $(i,j)$ and a reference beam $(i_R,j_R)$~\cite{zupancic2016ultra}.
As shown in Ext. Data Fig.~\ref{efig:fige2}, the SLM isolates the reference beam $(i_R,j_R)$ and the target beam $(i,j)$ out by applying local blazed-grating masks.
We extract the relative phase from their interference pattern in the Fourier plane (Ext. Data Fig.~\ref{efig:fige2}\textbf{c}).
The line-cut intensity (Ext. Data Fig.~\ref{efig:fige2}\textbf{d}) is fitted to $I(x_f)=I_0\left(1+\gamma \cos\left(2 \pi f~x_f + \phi_{ij} \right)\right)$, where $\phi_{ij}$ is the phase difference relative to the reference. The origin of the coordinate axis $x_f$ is kept fixed for different $(i,j)$ beams.

During the calibration, the laser frequency and the round-trip length of the second RIPA are actively stabilized to ensure fringe stability.
After calibration, the calibrated phase mask (Ext. Data Fig.~\ref{efig:fige2}\textbf{b}) is applied to compensate for wavefront distortions.
Given the $\sim 3~\mathrm{GHz}$ spectral bandwidth, chromatic dispersion is negligible. Consequently, the calibration remains valid and maintains high interference quality across the entire operating frequency range.

\subsection{Characterization of RIPA}
Here we present the methods used to characterize the RIPA in terms of its frequency-domain response (Fig.~\ref{fig:fig3}, static, data acquired with a camera), and its time-domain dynamics (Fig.~\ref{fig:fig4}, transient, data acquired with a scanning photodetector).

\subsubsection*{Frequency-domain (static) characterization}
We detail the methods used to characterize the frequency-domain response of RIPA shown in Fig.~\ref{fig:fig3}.

First, we calibrate the RF frequency and amplitude responses of the system. We vary the RF modulation frequency $\nu_k^{\mathrm{mod}}$ from 9 to 13~GHz and measure the power in the $+1$ sideband ($I$) after the EOM and filter. The RF amplitude response is similarly calibrated by varying the RF output amplitude $A_k$ across the 8-Bit AWG range (0-255); see Supplementary Information for the calibration curve. 

Leveraging the calibration data $I(\nu_k^{\mathrm{mod}},A_k)$, we generate highly uniform arbitrary patterns and characterize their quality.
We drive the EOM with a single RF frequency and sweep across the first RIPA's free spectral range ($\mathrm{FSR}_1\sim 3~\mathrm{GHz}$) over an $11\times 11$ grid to characterize the interference peaks within the first Brillouin zone. Given $R=\mathrm{FSR}_1/\mathrm{FSR}_2\approx24.6$, we modulate the frequencies as $\nu_{k=(i,j)}^{\mathrm{mod}}=(2\cdot i+j/N)\cdot \mathrm{FSR}_2$ to address a quasi-squared grid.
The RIPA-SLM routes these tones to the image plane in a near-orthogonal grid (with relative angle $\theta=\pi/2-\arctan(1/R)$). 
The resulting spacings of the ``grid'' are $d_x=\frac{f\lambda}{p}\frac{1}{N}=14.2~\mu\mathrm{m}\sim w_0^\prime$, and $d_y=\frac{f\lambda}{p}\frac{2}{R}=12.7~\mu\mathrm{m}\lesssim w_0^\prime$.
A careful choice of $R=\mathrm{FSR}_1/\mathrm{FSR}_2=N=N_x=N_y$ and $\nu_{k=(i,j)}^{\mathrm{mod}}=(i+j/N)\cdot \frac{1}{1+1/R^2}\cdot\mathrm{FSR}_2$ will generate an orthogonal square grid $d_x=d_y\sim w_0^\prime$, achieving the optimal frequency-to-spatial mapping for 2D site-resolved addressing (see Supplementary Information).

We image the RIPA-SLM focal plane using a Basler camera (a2A5060-15umBAS), which blocks the \review{$\lambda=785~\text{nm}$} locking-path light by a \review{$\lambda_c=780~\text{nm}$} MaxLine filter (Semrock, LL01-780).
Using OpenCV~\cite{opencv_library}, we extract interference peak positions and apply local 2D Gaussian fits to determine the $1/e^2$ waists ($w_x$, $w_y$) and peak intensities ($I$). 
This fitting is valid because the interference pattern locally resembles a 2D Gaussian (details in later sections).
Site-resolved heatmaps of (Fig.~\ref{fig:fig3}\textbf{g}-\textbf{h}) show standard deviations of $\sigma=3.1\%$ for $w_x$ and $\sigma=1.9\%$ for $w_y$, respectively.

Despite RF system calibration, the intensities $I$ are modulated by a Gaussian envelope which is the Fourier transform of the individual beams' common spatial mode.
By extracting and compensating for this envelope using the camera data, we achieve an intensity homogeneity of $\sigma=2.6\%$ (Fig.~\ref{fig:fig3}\textbf{f}).

To assess the system's suitability for optical site-addressing, we calibrate the crosstalk (Fig.~\ref{fig:fig3}\textbf{e}), defined as the Gaussian-weighted optical power encircled within a waist $w_0^\prime$ at a site located a distance $d$ from an addressed site.
The experimental results for an $(N_x,N_y)=(8,9)$ array align well with the theoretical model, accounting for measured power attenuation of 19.8\% (5.2\%) per $x(y)$ round trip; these values include out-coupling ratios of 6.8\% (3.0\%) per round trip.
This suggests that for a \review{$(N_x,N_y)=(80,80)$} array with a 1\% loss per round trip in both RIPAs (red curve, Fig.~\ref{fig:fig3}\textbf{e}), crosstalk would reach $10^{-3}$ at a separation of $6.5$ waists and $10^{-4}$ at a separation of $13.5$ waists. 

The multi-beam interference nature causes the second-largest interference peak to maintain an intensity of $\left(\frac{2}{3\pi}\right)^2\approx4.5\%$ relative to the main peak, regardless of the number of beams $N$. See Supplementary Information for the curves.
These significant interference ``tails'' drive the power-law scaling of crosstalk observed in Fig.~\ref{fig:fig3}\textbf{e}.
Conversely, this enables operation in a closely-packed regime where $d_{x,y}=\frac{f\lambda}{p}\frac{1}{N_{x,y}}$; at this grid spacing, each spot coincides with a null in the field of its neighbors.
We also report crosstalk scaling curves along the x- axis and the diagonal axis, see Supplementary Information for details.

\subsubsection*{Time-domain (transient) characterization}
\label{sec:method_scanning_pd}
We detail the methods used to characterize the nanosecond-scale time-domain dynamics of the RIPA-SLM using a scanning photodetector, as presented in Fig.~\ref{fig:fig4}.

The measurement principle of the scanning photodetector method relies on capturing repeatable time-resolved response of the RIPA focal plane at discrete spatial coordinates. 
By physically translating the photodetector across the target region, we reconstruct the full spatiotemporal evolution of the optical field. 

As shown in Ext. Data Fig.~\ref{efig:fige3}\textbf{a}, we place a $5~\mu\mathrm{m}$-diameter pinhole (Thorlabs P5K) in front of an avalanche photodiode (APD, Thorlabs APD130A2), which is mounted on a three-axis translation stage.
The detector is precisely positioned along the $z$-axis to coincide with the focal plane ($xOy$).
At each scanning coordinate, the control PC initiates a trigger to synchronize the AWG and an oscilloscope.
Once triggered, the AWG then generates the programmed RF pulse sequences to drive the EOM, while the oscilloscope records the APD voltage traces (Ext. Data Fig.~\ref{efig:fige3}\textbf{b}, inset) with 2~ns temporal resolution.
By systematically acquiring time traces over a 2D grid covering the first Brillouin zone with $5~\mu\mathrm{m}$ spatial resolution, we reconstruct the complete 3D spatiotemporal intensity distribution (Ext. Data Fig.~\ref{efig:fige3}\textbf{b}).

To demonstrate the versatility and reconfigurability of RIPA, we characterize the asynchronous operation of independent optical spots. We report \review{an extra dataset showing} the reconstructed dynamics of two spots following a 1D zigzag trajectory (Ext. Data Fig.~\ref{efig:fige3}\textbf{c}) exhibiting a characteristic switching time of 200~ns.

These results, along with those in Fig.~\ref{fig:fig4}, are obtained with a tapered amplifier (TA) placed between the EOM and the second-order filter. The observed dynamics include contributions from the RF system, the TA, the photodetector, and the round-trip propagation within the RIPAs.
The fact that we could still observe fast dynamics is attributed to the intrinsic high bandwidth of the TA, alleviating EOM power-handling limitations.
Numerical simulation of the RIPA pulsed rising response is shown in Ext. Data. Fig.~\ref{efig:fige4}\textbf{c}, revealing a quantitative agreement with the experimental result.

\subsection{Properties of RIPA round trips}
\review{Here, we describe the key properties of RIPA round trips that underlie the stability and performance of the system. Specifically, we operate in a ``half-confocal'' lens-guide mode, which suppresses the accumulation of optical imperfections.}

\subsubsection*{Trajectory stabilization with an MLA against optical imperfections}
In the absence of the MLA, the round trips are susceptible to imperfections such as mirror curvature, aberrations, astigmatism, and surface roughness.
\review{
The microlenses in the beam path form a lens-guide that re-images the beam onto itself provided that the input beam is mode-matched to the eigenmode of the lens-guide.
Optical imperfections perturb the wavefront and couple this mode to higher-order transverse eigenmodes, as described by a unitary transformation $\hat{U}_\mathrm{perturb}$ of the form $\hat{U}_\mathrm{perturb}=e^{-i\epsilon\hat{h}}\sim\mathbb{I}-i\epsilon\hat{h}$, where $\hat{h}=h_{ij}\ket{i}\bra{j}$ is the coupling matrix with no diagonal terms, and $\{\ket{i}\}$ are the transverse modes, with indices sorted by the transverse mode order (the first one being the $\text{TEM}_{00}$ mode).}

\review{Because these imperfections typically vary slowly on the scale of the MLA pitch, consecutive round trips impose nearly identical perturbations. The MLA, however, imparts an extra Gouy phase $\theta_\mathrm{gouy}$ to the higher-order eigenmodes, corresponding to another unitary transformation between round trips, $\hat{U}_\mathrm{gouy}=\mathrm{diag}(1,e^{i k\theta_\mathrm{gouy}},\cdots)$, where $k\in\mathbb{Z}^+$ denotes the transverse mode order. 
After $N$ round trips, the total evolution is thus $\hat{U}_\mathrm{tot}=(\hat{U}_\mathrm{perturb}\hat{U}_\mathrm{gouy})^N$.
Consequently, mode amplitudes generated by slowly varying perturbations acquire different phases on successive round trips and therefore interfere destructively rather than constructively: 
\[\hat{U}_\mathrm{tot}\sim(\mathbb{I}-i\epsilon\underbrace{\sum_{n=0}^{N-1}(\hat{U}_\mathrm{gouy})^{n}\hat{h}(\hat{U}_\mathrm{gouy}^\dagger)^{n}}_{\to0})(\hat{U}_\mathrm{gouy})^N+\mathcal{O}(\epsilon^2),\]
where the first-order term cancels because each matrix element contains a geometric sum of the form $\sum_{n=0}^{N-1}h_{ij} e^{i\theta_\mathrm{gouy}n\alpha}$, which vanishes for a uniform or slowly varying $\hat{h}$. 
}


Alternatively, following Ref.~\cite{sommer2016engineering}, the ray can be viewed as evolving under an adiabatically time-varying Hamiltonian induced by optical imperfections.
Rays initialized in an eigenstate of this Hamiltonian remain in an instantaneous eigenstate throughout this adiabatic evolution because of the finite energy gap arising from the MLA's transverse confinement.
In contrast, without the MLA, the 4f system is degenerate for all beams and lacks this stabilization property.

\subsubsection*{``Half confocal'' lens-guide mode}
The round-trip dynamics of a beam within the RIPA lens-guide comprises propagation over a distance $L_\mathrm{rt}/2$, a microlens with focal length $f_{\mathrm{MLA}}$, and another propagation over $L_\mathrm{rt}/2$.
In terms of the ABCD Matrix, we have
\begin{align}
\begin{split}
M &= \left(\begin{matrix}
1 & L_\mathrm{rt}/2\\0&1
\end{matrix}\right)\left(\begin{matrix}
1 & 0\\-\frac{1}{f_{\mathrm{MLA}}}&1
\end{matrix}\right)\left(\begin{matrix}
1 & L_\mathrm{rt}/2\\0&1
\end{matrix}\right)\\
&=\left(\begin{matrix}
1-\frac{L_\mathrm{rt}}{2f_{\mathrm{MLA}}} & L_\mathrm{rt}-\frac{L_\mathrm{rt}^2}{4f_{\mathrm{MLA}}}\\
-\frac{1}{f_{\mathrm{MLA}}}&1-\frac{L_\mathrm{rt}}{2f_{\mathrm{MLA}}}
\end{matrix}\right)\equiv\left(\begin{matrix}
A & B\\C&D
\end{matrix}\right).
\end{split}
\end{align}
The eigenmode Gaussian parameter $q_e$ satisfies $q_e=(Aq_e+B)/(Cq_e+D)$, which yields the solution $q_e=i\sqrt{L_\mathrm{rt}(f_{\mathrm{MLA}}-L_\mathrm{rt}/4)}$.
The resulting mode waist is $w_0=\sqrt{\lambda\,\mathrm{Im}(q)/\pi}$.
To achieve the maximum alignment stability against mode-mismatch, we operate at $L_\mathrm{rt}=2f_{\mathrm{MLA}}$ and $w_0=\sqrt{f_{\mathrm{MLA}}\lambda/\pi}\sim108~\mu\text{m}$. This configuration corresponds to a Gouy phase $\phi_\mathrm{gouy}=2\arctan(L_\mathrm{rt}/2/\mathrm{Im}(q))=2\arctan(1)=\pi/2$, which is exactly half of the Gouy phase obtained in a confocal cavity.
We denote this the ``half confocal'' regime.
Operating in this regime guarantees maximal mode overlap across the 2D phased array in the presence of input mode mismatch or aberration-induced distortions, maintaining high-quality interference.
Additionally, this condition minimizes the spot size on the MLA pupil ($w_\mathrm{MLA}=\sqrt{2} w_0=\sqrt{2\lambda f/\pi}$) similar to an optical cavity~\cite{jaffe2022understanding,shadmany2025cavity}, therby minimizing the clipping loss.

\subsubsection*{Ray tracing simulations}
Ray tracing simulations (Fig.~\ref{fig:fig1}\textbf{b}) were performed using an in-house, open-source package~\cite{Tim_optable} for non-paraxial propagation through spherical, aspheric, and microlens array optics.
The software’s GUI enables real-time manipulation of experimental variables, providing critical insights for optical alignment.

To simulate the 2D beam array interference pattern at the focal plane, we utilized an in-house, open-source wave propagation package~\cite{Tim_vipa_focus_simulation}. This allows for wave-optics analysis accounting for misalignment, phase errors, and time-dependent driving sequences.

\subsection{Properties of RIPA output}

\subsubsection*{Electric field in the focal plane}
\label{sec:e_field_in_focal}
\review{
The 2D phased array comprises a set of mutually coherent beamlets with an identical Gaussian profile $G(\mathbf{r},w_0)$ at positions $p(i\mathbf{e}_x+j\mathbf{e}_y)$. Here, $w_0$ is the waist of the Gaussian beam, $p$ is the array pitch, and $\mathbf{e}_x$ and $\mathbf{e}_y$ are unit vectors along the $x$ and $y$ directions, respectively. 
The phased array can be expressed mathematically as the \emph{convolution} ($\star$) of the Gaussian profile and a comb function. The electric field in the RIPA focal plane, given by the Fourier transform of this phased-array field, can be expressed as:
}
\begin{align}
\begin{split}
&E_\mathrm{focal}\left(k_x,k_y\right)=\mathcal{F}[E_\mathrm{array}(\mathbf{r})]\\
=&\mathcal{F}\left[G(\mathbf{r},w_0)\star\sum_{i=0,j=0}^{N_x-1,N_y-1}e^{\mathbb{i}(i\varphi_x+j\varphi_y)}\delta(\mathbf{r}-p(i\mathbf{e}_x+j\mathbf{e}_y))\right]\\
=&\mathcal{F}\left[G(\mathbf{r},w_0)\right]\cdot e^{-\mathbb{i}\phi_0}\\
&\cdot 
\frac{\sin\left(\frac{N_x}{2}\left(k_x p -\varphi_x\right)\right)}{\sin \left(\frac{1}{2}\left(k_x p -\varphi_x\right)\right)}
\cdot\frac{\sin\left(\frac{N_y}{2}\left(k_y p -\varphi_y\right)\right)}{\sin \left(\frac{1}{2}\left(k_y p -\varphi_y\right)\right)}
\end{split}
\end{align}
where $\delta(\cdot)$ is the Dirac delta function, $(i,j)$ index the beams, and $(N_x,N_y)=(8,9)$ is the size of the beam array. $\varphi_x$ and $\varphi_y$ are the phase steps between adjacent beams along the $x$ and $y$ directions, respectively, and $\phi_0$ is a global phase that does not affect the intensity distribution.

A lens maps the field component at spatial frequencies $k_x,k_y$ to the image-plane coordinates $x_f = \frac{\lambda f}{2\pi}k_x,\ y_f = \frac{\lambda f}{2\pi}k_y$, yielding the intensity:
\begin{align}
\begin{split}
&\left|E_\mathrm{focal}\left(x_f,y_f\right)\right|^2\\
=&\left|G\left(x_f,y_f,\frac{f\lambda}{\pi w_0}\right)\right|^2\cdot 
\frac{\sin^2\left(N_x\pi \frac{x_f-x_0}{L}\right)}{\sin^2\left(\pi \frac{x_f-x_0}{L}\right)}
\frac{\sin^2\left(N_y\pi \frac{y_f-y_0}{L}\right)}{\sin^2\left(\pi \frac{y_f-y_0}{L}\right)}
\end{split}
\label{equ:grating_intensity}
\end{align}
Here, $L=\frac{f\lambda}{p}$ is the Brillouin zone extent, and $(x_0,y_0)=(L\frac{\varphi_x}{2\pi},L\frac{\varphi_y}{2\pi})$ defines the interference peak center.
Equation.~\ref{equ:grating_intensity} reflects a standard grating distribution arising from multi-beam interference, modulated by a Gaussian envelope $\left|G\left(x_f,y_f,\frac{f\lambda}{\pi w_0}\right)\right|^2$ with waist $w_{\mathrm{env}}=\frac{f\lambda}{\pi w_0}$, which is the Fourier transform of the common Gaussian mode (with waist $w_0$) that each beam within the 2D phased array shares.

In the vicinity of the interference peak, the \textit{grating function} $g(x)\equiv\frac{\sin^2\left(N\pi x/L\right)}{\sin^2\left(\pi x/L\right)}$ can be approximated by a Gaussian $g(x)=N^2\exp\left(-\alpha(\frac{x}{L})^2\right)+o(x^4)$, where $\alpha=\frac{\pi(N^2-1)}{3}$.
The equivalent Gaussian waist is $w_0^\prime=L\sqrt{\frac{2}{\alpha}}=\frac{f\lambda}{\pi p}\sqrt{\frac{6}{N^2-1}}$. 
As detailed in the Supplementary Information, the deviation from this approximation becomes significant only beyond one waist $w_0^\prime$ away from the peak.
Under typical operating conditions ($p\gg w_0,\ N\gg1$), the focal-plane intensity is \emph{grating-limited}: the peak width and spacing are governed by the pitch $p$, while the Gaussian envelope remains quasi-uniform across individual interference peaks.

\subsubsection*{Number of Brillouin zones and power concentration}
The ratio between the MLA pitch $p$ and the RIPA beam waist $w_0$ determines the effective number of Brillouin zones, $M=w_\mathrm{env}/L=p/(\pi w_0)$.
To minimize round-trip clipping losses at the MLA pupil, the RIPA typically operates in the regime $p\gg w_0$, yielding $M\gg1$.
While this enables parallel spatial addressing over a $\sim M\times M$ grid, it conversely suffers from the \review{low power efficiency within the first Brillouin zone}.

Optical power can be concentrated within the first Brillouin zone by increasing $w_0$ relative to $p$ using an MLA-telescope (focal lengths $f_1,f_2$; separation $d=f_1+f_2$) at the array output, effectively forming a local telescope for each individual beam.
The resulting magnification increases clipping loss, which remains acceptable because the beam undergoes only a single pass.

\subsubsection*{Off-focal intensity distribution}
Numerical simulations confirm that the intensity distribution beyond the focal plane approximates that of a Gaussian beam (Ext. Data Fig.~\ref{efig:fige4}\textbf{d},\textbf{f}).
This behavior can be understood intuitively by treating the 2D phased array as a ``distributed'' large beam with a tilted wavefront and a quasi-continuous amplitude profile.
Simulations at discrete frequencies demonstrate that changing the phase difference $\varphi_x$ shifts the beam ``waist'' along the $x$ direction, while its $z$-position remains fixed.
The intensity asymmetry along the $x$ axis in Ext. Data Fig.~\ref{efig:fige4}\textbf{d},\textbf{f} results from monotonic power decay due to round-trip losses.

\subsubsection*{Tolerance to misalignment}
System misalignment, stemming from mode mismatch or optical aberrations, is analyzed by decomposing each beam into the local Hermite-Gaussian basis $\ket{lm}=\mathrm{TEM}_{lm}$ on each microlens.
In an ideal case, all beams share the fundamental Gaussian mode $\ket{00}$.
Tilt or positional misalignments primarily excite the lowest-order odd modes $\ket{10}$ or $\ket{01}$, while a mode-size mismatch primarily excites even modes such as $\ket{20}$.
These modes possess distinct Gouy phases $\psi(z)=(l+m+1)\arctan\left(\frac{z}{z_R}\right)$ and therefore focus at different transverse positions relative to the $\ket{00}$ mode in the focal plane.

\review{Because the Hermite function associated with the $\ket{10}$ or $\ket{01}$ modes vanishes at the center of the Brillouin zone and remains small near it, these modes contribute much less power than the fundamental $\ket{00}$ mode within the first Brillouin zone.}
To quantify this suppression, we integrate the field intensity over the first Brillouin zone, defining the suppression factor 
$S=\frac{\int_{1^\mathrm{st}\,\text{Brillouin zone}}|E_{00}|^2\mathrm{d}x_f\mathrm{d}y_f}{\int_{1^\mathrm{st}\,\text{Brillouin zone}}|E_{01}|^2\mathrm{d}x_f\mathrm{d}y_f}$.
In our system, the suppression $S$ ranges between $10.7$ and $249$ for different input laser frequencies, with an average value of $66.4$.

\subsubsection*{Lensing effect compared with AOD}
We evaluate the RIPA-equivalent acoustic lensing effect as in AODs~\cite{zhang2025leveragingerasureerrorslogical,dickson1972optical}. 
By calculating the focal shift $\delta z$ induced by frequency chirping in both RIPA and AOD, we derive the ratio (See the Supplementary Information for a detailed derivation):
\begin{align}
    \left|\frac{\delta z_{\mathrm{RIPA}}}{\delta z_{\mathrm{AOD}}}\right|=\frac{1}{\sqrt{6}}\frac{\tau_{\mathrm{RIPA}}}{\tau_{\mathrm{AOD}}}
\end{align}
under identical sweeping speeds and focal waist size $w_0^\prime$ in the focal plane.
Here $\tau_{\mathrm{RIPA}}=\frac{NL_{rt}}{c}$ and $\tau_{\mathrm{AOD}}=\frac{w_0}{v_s}$ denote the characteristic propagation times in RIPA and AOD, respectively. 
Consequently, our current device exhibits a suppressed lensing effect (0.027$\times$) relative to an AOD with a $\tau_{\mathrm{AOD}}\sim 1~\mu\mathrm{s}$ switching time.
The numerical simulation results presented in Ext. Data Fig.~\ref{efig:fige4}\textbf{e} shows the lensing trajectory under linear frequency chirping of various durations ($T=$ 200, 500, and 2000~ns), confirming the fast and continuous steering of the addressing beam with a tolerable lensing effect.

\subsection{Scaling up the RIPA}

\subsubsection*{Power efficiency and loss budget}
\begin{table}
    \label{tab:tabe1}
    \centering
    \caption{\textbf{Power efficiency analysis of the RIPA.}
    Internal loss per round trip comprises absorption, clipping, and scattering into higher-order modes.}
    \begin{tabular}{lll}
        \toprule
        \textbf{Quantity}                                      & \textbf{Symbol}                                                            & \textbf{Value (\%)}               \\
        \midrule
        $1^{\mathrm{st}}$ RIPA RT loss & $A_1=\kappa_1+l_1$                                                & 5.1(9)  \\ 
        \quad - Out-coupling                 & $\kappa_1$                                                        & 2.9(7) \\ 
        \quad - Internal loss                 & $l_1$                                                             & 2.2(1.1) \\ 
        $1^{\mathrm{st}}$ RIPA efficiency             & $\eta_1=\kappa_1\cdot\sum_{j=0}^{N_y-1} (1-A_1)^j$                   & 21.9(2) \\[3pt]
        Relay system efficiency                       & $\eta_{\mathrm{rel}}$                                             & 70.6(4) \\[3pt]
        $2^{\mathrm{nd}}$ RIPA RT loss & $A_2=\kappa_{2,\mathrm{lock}}+\kappa_2+l_2$                       & 19.8(1.8) \\
        \quad - Locking path          & $\kappa_{2,\mathrm{lock}}$                                        & 9.7(9) \\ 
        \quad - Out-coupling                 & $\kappa_{2}$                                                      & 6.8(7) \\ 
        \quad - Internal loss                 & $l_2$                                                             & 3.3(2.2) \\ 
        $2^{\mathrm{nd}}$ RIPA efficiency             & $\eta_2=\kappa_2\cdot\sum_{i=0}^{N_x-1} (1-A_2)^i$                   & 28.3(5) \\[3pt] 
        SLM+imaging efficiency               & $\eta_{\mathrm{im}}$                                              & 85(5) \\[3pt]
        \midrule
        \textbf{Total system efficiency}                       & $\eta=\eta_1\cdot \eta_{\mathrm{rel}}\cdot \eta_2\cdot \eta_{im}$ & \textbf{3.7(1)} \\
        \bottomrule
    \end{tabular}
\end{table}

The power-efficiency analysis is detailed in Ext. Data Table~\ref{tab:tabe1}.
The round-trip losses $A_1$ and $A_2$ are extracted by fitting the intensity distribution of the output 2D beam array to an exponential decay model.
These losses comprise out-coupling and internal losses; notably, the second stage incorporates a locking-path beamsplitter ($\kappa_{2,\mathrm{lock}}$) that introduces additional loss.
The measured total system efficiency of 3.7(1)\% is primarily limited by the out-coupling ratio (see discussion in the following section).

The fundamental scalability of the RIPA is governed by internal losses (if the internal loss is high, the beams cannot make many round trips in the RIPA). Transitioning to a planar two-mirror configuration would eliminate the locking path, substantially reducing intrinsic losses.
Currently, the round-trip loss is 2--3\%, comprising mirror losses ($\sim 1\%$, 1$\times$ Thorlabs BB1-E03, 3$\times$ Thorlabs UM10-45A), lens surface reflections ($\sim1\%$, 8$\times$ V-coated surfaces), and MLA reflections ($\sim1\%$, Edmund Optics \#21-155). Replacing the MLA with a high-reflectivity micro-mirror array (MMA) further reduce the surface reflection loss (see Supplementary Information).
We calculate the clipping loss at the MLA to be $\exp\left(-\frac{2(d/2)^2}{(\sqrt{2}w_0)^2}\right)=0.005\%$ which is negligible, where $d=500~\mu$m is the MLA pupil size. However, due to imperfect mode-matching or aberration-induced oscillations, the higher-order modes may experience larger clipping loss at the MLA. \review{On the other hand, this also serves as an effective mechanism to damp higher-order modes.}

\subsubsection*{Power handling}
\review{Each beamlet in our phased array has a mode-field diameter (MFD) of $2\sqrt{2}w_0=305~\mu\mathrm{m}$ at the MLA aperture. The MLA substrate (Edmund Optics \#21-155) is fused silica and has negligible optical absorption; accordingly, the power handling is ultimately limited by the anti-reflective (AR) coating, which in CW operation typically has a damage threshold on the order of $\mathrm{kW\,cm^{-1}}$. This corresponds to a maximum input power at the microlens array of approximately 30~W. With a micromirror-array design and a high-reflectivity (HR) coating, the system could potentially handle substantially higher input power.}

\review{We also estimate the power-handling capability of the other optics. The free-space second-order optical filter is essentially a coupled Fabry--Perot cavity pair, with estimated parameters of waist $w=66~\mu\mathrm{m}$, a damage threshold of $10~\mathrm{kW\,cm^{-1}}$ estimated from Thorlabs BB1-E03 mirrors, and finesse $\mathcal{F}\sim20$. It can therefore support up to $\sim20~\mathrm{W}$ of output power. The \review{LCoS-}SLM used here (Meadowlark E-series) was chosen for demonstration purposes, whereas higher-power models can handle hundreds of watts, as exemplified by the Hamamatsu X15213-02L.}

\subsubsection*{Footprint scaling}
The system footprint is primarily determined by the round-trip length of the second (and long) RIPA. To maintain a similar modulation bandwidth (i.e. frequency resolution of the long RIPA $f_{\mathrm{res}}=c/(N_x L_{\mathrm{r,t2}})\sim \text{MHz}$), the round-trip length of the long RIPA in a scaled-up device (\review{with increased $N_x, N_y$}) will be, actually, \textit{shorter} than the current design.
Under scaling, the attainable EOM aggregation bandwidth (tens of GHz) becomes the primary constraint rather than the physical footprint, \review{which can be solved by stacking multiple EOM modules (see Supplementary Information)}. Numerical analysis indicates that a \review{$80\times80$} RIPA-SLM with $1~\text{MHz}$ per-pixel bandwidth operates at a sweet point, with an aggregation bandwidth of \review{$6.8~\text{GHz}$} and a meter-scale footprint of \review{$L_{\mathrm{r,t2}}\sim 3.7~\text{m}$}.
\review{We also proposed a \emph{lens-free} design which would shrink the footprint of an $80\times80$ RIPA to about $10~\text{cm}\times 10~\text{cm}$. See the Supplementary Information for details. }

To conclude, adopting a two-mirror geometry, MMAs, and nano-textured optics should reduce round-trip loss to $<1\%$.
This improvement would support hundreds of round trips, enabling \review{$80\times80$} beam arrays in the near term, \review{with power handling capability up to 30~W}.

\subsubsection*{Tradeoff between power efficiency and linewidth}
\label{sec:tradeoff_eff_lw}
Increasing the power out-coupling ratio per round trip improves the RIPA system’s power efficiency. However, the resulting intensity decay in successive round trips degrades the interference pattern as it reduces the effective number of interfering beams and thereby broadens the linewidth.
This establishes an inherent tradeoff between power efficiency and interference linewidth.
We numerically simulate the $1/e^2$ waist ($w$) of the interference peak as a function of power out-coupling ratio ($\kappa$) with a certain internal loss ($l$).
We define the linewidth broadening factor $R_w\equiv w/\lim_{\kappa\to 0,l_\to0} w$ as the ratio of $w$ to the diffraction-limited waist attained when all output beams have uniform intensity.

Ext. Data Fig.~\ref{efig:fige5} illustrates this tradeoff for the first ($N_y=9$, internal loss $l_1=2.2\%$), the second RIPA ($N_x=8$, internal loss $l_2+\kappa_{2,\mathrm{lock}}=13\%$), and the \review{single stage of} the proposed \review{$80\times80$} RIPA \review{($N_x=N_y=80$, internal loss $l_{80}=1\%$)}.
Experimentally, we operate \review{the current device} at $\kappa_1=2.9\%$ and $\kappa_2=6.8\%$ (gray dashed lines).
Depending on the application, the system can be optimized for either power efficiency or interference quality.
This tradeoff is remarkably favorable: with sufficiently low internal loss, one can utilize $>80\%$ of the optical power without significantly interference pattern degradation ($<20\%$ linewidth increase).

\subsection{\review{Building a neutral atom processor around the RIPA-SLM}}
\review{The frequency-to-position mapping of RIPA provides a natural solution for scalable local control in neutral atom arrays. This mapping supports two complementary modes of operation, as illustrated in Ext. Data Fig.~\ref{efig:fige6} \textbf{first,} static addressing: the RIPA-SLM is used to address selected atomic qubits at fixed locations for single-qubit internal-state manipulation (Ext. Data Fig~\ref{efig:fige6}\textbf{a}); \textbf{second,} reconfigurable moving: the RIPA-SLM dynamically produces moving tweezers for coherent atom transport and rearrangement (Ext. Data Fig~\ref{efig:fige6}\textbf{b-c}). Thus, the RIPA-SLM architecture combines local control and reconfiguration in a single frequency-multiplexed optical platform.}

\review{Because of its frequency variation across the image plane, the RIPA-SLM is not intended for \emph{direct} resonant single-photon operations on individual atoms. Instead, it can address individual atoms through operations that are weakly dependent on the single-photon detuning, including dispersive light shifts, Raman transitions, and far-detuned trapping potentials.
In these schemes, the single-photon detuning is typically from 200~GHz to a few THz, so the GHz-scale frequency span required to cover the field of view produces only a $\sim 10^{-3}$ change in the single-photon detuning.
The residual site-to-site variation can be mitigated by site-resolved calibration, or robust control protocols such as composite pulses and pulse shaping.
However, we also note that the multi-GHz spectral bandwidth is almost certainly too large for magic-wavelength trapping, which is used to maintain maximum coherence on long-lived optical transitions.
}

\review{Here we exemplify how to build a neutral atom processor around the RIPA-SLM by demonstrating its use for interfacing with hyperfine-encoded $^{87}\mathrm{Rb}$ atoms.}

\subsubsection*{Internal state manipulation}
\review{
\paragraph{Scalar AC-Stark shift. }
When an atom interacts with a single-frequency laser beam in the dispersive limit, the lowest-order effect can be a local scalar AC-Stark shift. 
This shift can lead to energy shifts on specific levels for shelving~\cite{chiu2025continuous}, local resonance tuning~\cite{wang2015coherent}, and Floquet state engineering~\cite{jia2018strongly}. 
Our RIPA-SLM can generate individually addressing beams for each qubit and perform all these operations at wish in an asynchronous manner. 
}

\review{
\paragraph{Vector AC-Stark shift. }
Similarly, by controlling the polarization or choosing wisely the quantization axis, the output of the RIPA-SLM can also generate 
vector AC-Stark shifts which offer programmable individual-qubit phase rotations $R_z(\theta)$. 
Because the addressing position is set by optical frequency, many such operations can be applied in parallel by synthesizing multiple frequency tones with independently programmed amplitudes and durations.
}

\review{
\paragraph{Two-photon Raman transition. }
The RIPA-SLM is a spectrometer with frequency periodicity $\mathrm{FSR_1}$. If $\mathrm{FSR_1}$ is chosen to be an integer fraction of the qubit energy separation, for example, the $^{87}\mathrm{Rb}$ ground-state hyperfine splitting $\mathrm{HFS}=6.835~\mathrm{GHz}$, then two optical frequency tones $\nu$ and $\nu+\mathrm{HFS}$ will be routed to the same position in the image plane, and thus address the same atom.
These two tones can drive a two-photon Raman transition between the hyperfine atomic states, thereby realizing single qubit $R_x(\theta)$ rotations for hyperfine-encoded qubits. 
Different choices of $\nu$ select different atoms in the image plane, and frequency-multiplexing enables site-resolved Raman control. The relative optical phase $\varphi$ between the two Raman tones sets the rotation axis in the equatorial plane of the Bloch sphere.
Thus, by controlling the amplitudes, duration, and relative phase of the two tones, the system can perform arbitrary site-resolved $R_\varphi(\theta)$ rotations and, together with local $R_z$ control, arbitrary single-qubit gates.
}

\review{
\paragraph{Resonant driving}
Single-photon resonant driving is a common requirement for imaging, in-situ cooling, single-photon Rydberg excitation, and related operations. In the RIPA-SLM architecture, site-selective single-photon resonant driving can be achieved by combining a global resonant excitation beam with scalar AC-Stark-shift local addressing from the RIPA-SLM to \emph{shield} untargeted atoms, as discussed above.
Another solution we envision here is to generate local drive fields using a pair of RIPA-SLMs with complementary frequency dispersion across the field of view (such that their sum frequency is spatially uniform), and to combine them by per-site sum-frequency generation (SFG), potentially assisted by a cavity array to enhance the conversion efficiency.
}

\subsubsection*{Coherent transport of atoms}
\review{
The same frequency-to-space mapping also enables coherent atom transport. Instead of switching the tones of discrete frequencies, one can continuously ramp the optical frequency to continuously steer the corresponding laser beam. The corresponding addressing spot then moves smoothly across the image plane, forming a moving optical tweezer. Atoms can be captured, transported, and released by programming the frequency trajectory and intensity envelope of each tone. 
}

\review{
To extend this motion to two dimensions, the RIPA output can be split into two independently controlled transport channels, namely \emph{row} and \emph{column} channels, as shown in Ext. Data Fig.~\ref{efig:fige6}\textbf{b}. The \emph{row} channel provides continuous sweeping motion along the horizontal direction, while the \emph{column} channel, rotated by $90^\circ$ in the image plane, provides continuous motion along the vertical direction.
After recombination and focusing onto the atom plane, the two channels generate orthogonal families of moving traps.
As illustrated in Ext. Data Fig.~\ref{efig:fige6}\textbf{c}, atoms can therefore be transported independently along rows and columns and transferred between the two channels at their intersections.
This handoff mechanism allows asynchronous reconfiguration of an atomic register, in which different atoms move along different trajectories and at different times rather than following a globally synchronized lattice translation or deformation.
}

\subsubsection*{Interference-resilient arbitrary trap patterning via frequency encoding}

\review{In many quantum gas experiments~\cite{bakr2009quantum,xu2025neutral} and atom array experiments~\cite{bluvstein2022quantum,bluvstein2024logical,bluvstein2026fault}, unwanted interference between nominally independent trapping fields is a common source of heating and loss. This issue becomes particularly severe when the optical potentials are dynamically reconfigured, as residual interference can produce time-dependent intensity corrugations near the motional response of the atoms.}

\review{For example, in AOD-based transport, the nonlinear acoustic response to multiple RF tones can generate intermodulation, producing spurious frequency components. When the modulation frequencies are not evenly spaced as a comb~\cite{bluvstein2022quantum,bluvstein2026fault}, these intermodulation products lead to time-varying interference, which can induce additional atom loss or heating. As a result, AOD-based transport is practically restricted to array stretching, compression, and translation, where the frequencies remain evenly spaced.}

\review{Our approach avoids this failure mode at the encoding level. Unlike LCoS- and photonic-integrated-circuit (PIC)-based SLMs~\cite{tara2026complex}, where the entire multi-trap potential shares a common optical frequency, our system assigns different traps to distinct spectral components. Within each trap, the constituent optical fields remain mutually phase-stabilized and frequency-degenerate; across traps, however, they are separated in frequency. As a result, residual optical or electrical crosstalk between neighboring channels produces beat notes of at least 16 MHz (RIPA's spectral resolution), far above the atomic motional response. This frequency encoding scheme is the key ingredient enabling high-quality pattern generation, as shown in Fig.~\ref{fig:fig3}\textbf{c}.
We note that the intermodulation in our system, which potentially arises from the RF amplifier and EOM, is substantially weaker than in an acoustic crystal.}

\clearpage

\end{document}